\documentclass[aps,preprint]{revtex4}
\usepackage{graphicx}
\usepackage{psfrag}
\usepackage{amssymb}

\begin{document}

%		DEFINITIONS FOR TEX
%
%%%%%%%%%%%%%%%%%%%%%%%%%%%%%%%%%%%%%%%%%%%%%%%%%%%%%%%%%%%%%%%
%
%
%\def\e{\'e}
%\def\ee{\`e}
%%%%%%%%%%%%%%%%%%%DEFINITIONS%%%%%%%%%%%%%%%%%%%%%%%%%%%%%%%%%
%
\def\oti{{\otimes}}
\def\lb{ \left[ }
\def\rb{ \right]  }
\def\tilde{\widetilde}
\def\bar{\overline}
\def\hat{\widehat}
\def\*{\star}
\def\[{\left[}
\def\]{\right]}
\def\({\left(}		\def\BL{\Bigr(}
\def\){\right)}		\def\BR{\Bigr)}
	\def\BBL{\lb}
	\def\BBR{\rb}
%
%%%%%%%%%%%%%%%%%%%%%%%%%%%%%%%%%%%%%%%%%%%%%%%%%%%%%%%%%%%%%%%
%
\def\zb{{\bar{z} }}
\def\zbar{{\bar{z} }}
\def\frac#1#2{{#1 \over #2}}
\def\inv#1{{1 \over #1}}
\def\half{{1 \over 2}}
\def\d{\partial}
\def\der#1{{\partial \over \partial #1}}
\def\dd#1#2{{\partial #1 \over \partial #2}}
\def\vev#1{\langle #1 \rangle}
\def\ket#1{ | #1 \rangle}
\def\rvac{\hbox{$\vert 0\rangle$}}
\def\lvac{\hbox{$\langle 0 \vert $}}
\def\2pi{\hbox{$2\pi i$}}
\def\e#1{{\rm e}^{^{\textstyle #1}}}
\def\grad#1{\,\nabla\!_{{#1}}\,}
\def\dsl{\raise.15ex\hbox{/}\kern-.57em\partial}
\def\Dsl{\,\raise.15ex\hbox{/}\mkern-.13.5mu D}
%
%%%%%%%%%%%%%%%%%%%%GREEK LETTERS%%%%%%%%%%%%%%%%%%%%%%%%%%%%%%
%
%\def\th{\theta}		\def\Th{\Theta}
\def\ga{\gamma}		\def\Ga{\Gamma}
\def\be{\beta}
\def\al{\alpha}
\def\ep{\epsilon}
\def\vep{\varepsilon}
\def\la{\lambda}	\def\La{\Lambda}
\def\de{\delta}		\def\De{\Delta}
\def\om{\omega}		\def\Om{\Omega}
\def\sig{\sigma}	\def\Sig{\Sigma}
\def\vphi{\varphi}
%
%%%%%%%%%%%%%%%%%%%CALIGRAPHIC LETTERS%%%%%%%%%%%%%%%%%%%%%%%%%
%
\def\CA{{\cal A}}	\def\CB{{\cal B}}	\def\CC{{\cal C}}
\def\CD{{\cal D}}	\def\CE{{\cal E}}	\def\CF{{\cal F}}
\def\CG{{\cal G}}	\def\CH{{\cal H}}	\def\CI{{\cal J}}
\def\CJ{{\cal J}}	\def\CK{{\cal K}}	\def\CL{{\cal L}}
\def\CM{{\cal M}}	\def\CN{{\cal N}}	\def\CO{{\cal O}}
\def\CP{{\cal P}}	\def\CQ{{\cal Q}}	\def\CR{{\cal R}}
\def\CS{{\cal S}}	\def\CT{{\cal T}}	\def\CU{{\cal U}}
\def\CV{{\cal V}}	\def\CW{{\cal W}}	\def\CX{{\cal X}}
\def\CY{{\cal Y}}	\def\CZ{{\cal Z}}

\def\rvac{\hbox{$\vert 0\rangle$}}
\def\lvac{\hbox{$\langle 0 \vert $}}
\def\comm#1#2{ \BBL\ #1\ ,\ #2 \BBR }
\def\2pi{\hbox{$2\pi i$}}
\def\e#1{{\rm e}^{^{\textstyle #1}}}
\def\grad#1{\,\nabla\!_{{#1}}\,}
\def\dsl{\raise.15ex\hbox{/}\kern-.57em\partial}
\def\Dsl{\,\raise.15ex\hbox{/}\mkern-.13.5mu D}
%
%%%%%%%%%%%%%%%%%%%%GREEK LETTERS%%%%%%%%%%%%%%%%%%%%%%%%%%%%%%
%
%%%%%%%%%%%%%%% MATH CHARACTERS %%%%%%%%%%%%%%%%%%%%%%%%%%%%
%
\font\numbers=cmss12
%\font\numbers=cmu10 scaled\magstep1
\font\upright=cmu10 scaled\magstep1
\def\stroke{\vrule height8pt width0.4pt depth-0.1pt}
\def\topfleck{\vrule height8pt width0.5pt depth-5.9pt}
\def\botfleck{\vrule height2pt width0.5pt depth0.1pt}
\def\Zmath{\vcenter{\hbox{\numbers\rlap{\rlap{Z}\kern
0.8pt\topfleck}\kern 2.2pt
                   \rlap Z\kern 6pt\botfleck\kern 1pt}}}
\def\Qmath{\vcenter{\hbox{\upright\rlap{\rlap{Q}\kern
                   3.8pt\stroke}\phantom{Q}}}}
\def\Nmath{\vcenter{\hbox{\upright\rlap{I}\kern 1.7pt N}}}
\def\Cmath{\vcenter{\hbox{\upright\rlap{\rlap{C}\kern
                   3.8pt\stroke}\phantom{C}}}}
\def\Rmath{\vcenter{\hbox{\upright\rlap{I}\kern 1.7pt R}}}
\def\Z{\ifmmode\Zmath\else$\Zmath$\fi}
\def\Q{\ifmmode\Qmath\else$\Qmath$\fi}
\def\N{\ifmmode\Nmath\else$\Nmath$\fi}
\def\C{\ifmmode\Cmath\else$\Cmath$\fi}
\def\R{\ifmmode\Rmath\else$\Rmath$\fi}
%%%%%%%%%%%%%%%%%%%%%%%%%%%%%%%%%%%%%%%%%%%%%%%%%%%%%%%%%%%%%%%%%
 %%%%%%%%%%%%%%%%%% END OF DEFINITIONS %%%%%%%%%%%%%%%%%%%%%%
 %%%%%%%%%%%%%%%%%%%%%%%%%%%%%%%%%%%%%%%%%%%%%%%%%

\def\barray{\begin{eqnarray}}
\def\earray{\end{eqnarray}}
\def\beq{\begin{equation}}
\def\eeq{\end{equation}}

\def\no{\noindent}

\title{Quasi-particle re-summation and integral gap equation
  in thermal field theory}
\author{Andr\'e  LeClair}
\affiliation{Newman Laboratory, Cornell University, Ithaca, NY} 
\date{August, 2004}

\bigskip\bigskip\bigskip\bigskip

\begin{abstract}

A new approach to quantum field theory at finite temperature and  density
in arbitrary space-time dimension $D$
is developed.  We focus mainly on relativistic theories, but the approach 
applies to non-relativistic ones as well. 
 In this quasi-particle re-summation, the free energy
takes the free-field form but with the one-particle energy
$\omega ( \vec{k} )$ replaced by $\vep (\vec{k} )$, the
latter satisfying a temperature-dependent integral equation
with kernel related to a zero temperature 
 form-factor of the trace of stress-energy
tensor.  For 2D integrable theories the approach reduces to 
 the thermodynamic Bethe ansatz.   For relativistic theories,  
a thermal c-function $C_{\rm qs} (T)$ is defined for 
any $D$ based on the coefficient of the black body radiation
formula. Thermodynamical constraints on it's flow are 
presented, showing that it can violate a ``c-theorem'' even 
in 2D.      At a  fixed
point $C_{\rm qs}$ is a function of thermal gap parameters 
which generalizes Roger's dilogarithm to higher dimensions.
This points to a strategy for classifying rational theories
based on ``polylogarithmic ladders'' in mathematics, and many
examples are worked out.   
Other applications are discussed,  including the free energy of
anyons in 2D and 3D, phase transitions with a chemical potential, 
and the equation of state for cosmological dark energy. For the latter
we show that $-1< w < -1/3$.

\end{abstract}

%\pacs{{\bf ??????}}
%\pacs{{\bf CHECK} 11.10.Hi, 11.55.Ds, 75.10.Jm}

\maketitle

\def\om#1{\omega_{#1}}

\def\Tr{\rm Tr} 
\def\free{\CF} 
\def\xvec{{\bf x}}
\def\kvec{{\bf k}}

\section{Introduction}

Quantum field theory at finite temperature and density 
was developed long ago mainly to deal with condensed matter
systems at low temperatures.  The subject was later taken 
up in the relativistic context where the interest was mainly
in high temperatures\cite{dellac,kapusta}, with the aim for example
to study hot QCD\cite{Wilczek}.    In both of these
contexts the main computational tool is the beautiful but sometimes
awkward sums over Matsubara frequencies in perturbation theory.  
It was later realized that to capture some important physics, 
such as the gluon relaxation rate, a re-summation scheme was necessary,
beyond the re-summation of ``ring diagrams'', 
at least to re-sum the hard thermal loops\cite{Braaten}.  
In spite of this important progress, finite temperature field theory
remains difficult,  especially when dealing with phase transitions,
and it is worthwhile to explore alternative methods.  
In the condensed matter context other kinds of self-consistent
re-summations are often used, for example in the finite
temperature gap equation of the BCS theory of superconductivity. 
The main purpose of this paper is to develop a new  approach to 
finite temperature field theory in any dimension $D$ 
in a similar but essentially different
spirit.  Because the object we are dealing with, i.e. a quantum field
functional integral in finite size geometry,  is so general, 
there are many possible applications;  we will touch on 
a number of them, such as central charges at fixed points,
classical statistical mechanics of Ising-like models in any
dimension, anyons and cosmology.

\def\kvec{{\bf k}}
\def\cqstat{{C_{\rm qs}}}
\def\cfp{{c_{\rm qs}}}

In sections II and III we introduce the usual quantities of
quantum statistical mechanics, the free energy,  pressure, entropy, etc.
and express them as one-point functions of the stress-energy tensor.   
It is meaningful to express everything in terms of a temperature
dependent c-function $\cqstat (T)$ which  at a  renormalization group 
(RG) fixed point
becomes the coefficient in the black body formula.  
  (The subscript ${}_{\rm qs}$ stands for
{\it quantum statistical} and is meant emphasize the distinction with
other c-functions.)
We show that in 4D,  this thermal c-function 
is different from other  c-functions  based on conformal
anomalies\cite{Cardyc4}. We also argue that thermodynamic principles,
such as positivity of the entropy,  
 do not lead 
to any  ``c-theorem's'' for $\cqstat$.  

$\cqstat (T)$ is expected to be important for 
equations of state in the thermal universe.   In section IV 
we relate the dark energy parameter ${ w}$ to a temperature
derivative of $\log \cqstat$.   Using the entropy bounds on 
$\cqstat$ we argue that for negative pressure $-1 < { w} < -1/3$.

The main workings of this paper begin in section V.   The idea is 
to assume that the partition function can be approximated by 
the standard free field form, but with the one particle energies
$\omega_\kvec$ replaced by a quasi-particle energy $\vep (\kvec )$.
A saddle point approximation to the partition function then leads
to an integral gap equation satisfied by $\vep$. The kernel in 
the integral equation is then related to zero temperature 
two-particle form factor of the trace of the stress-energy tensor
$T^\mu_\mu$.   
This approach is modeled after the thermodynamic Bethe ansatz
 (TBA)\cite{Yang,ZamoTBA,LecMuss}, which  is exact for
the 2D integrable quantum field theories.    For the non-integrable
theories in D dimensions we don't expect the results to be exact, 
but rather to be a  potentially useful approximation. 
In section VI we check that to lowest
order our formalism gives the same result as the two-loop Matsubara
formalism for the scalar $\phi^4$ theory.

At an RG fixed point, $\cqstat$ becomes a constant $\cfp$  which is
an important characteristic of the conformal field theory.
There are no known methods for computing $\cfp$ for interacting
fixed points in dimension $D>2$.  As a step in this direction, 
in section VII we describe how to compute $\cfp$ in  our
formalism.  The basic mechanism, which we believe to be general, is that 
at a fixed point the quasi-particle energy $\vep$ develops a gap $\Delta$  that
depends only on the temperature $\Delta = g T$.   The gap parameter
$g$ can be computed from an algebraic limit of the
 gap equation, and $\cfp$ is
a function only of $g$ which is a higher dimensional generalization 
of Roger's dilogarithm.  In this way a fixed point theory is deconstructed
into the thermal gap data, which depends only on {\it interaction
parameters} that are an integral of the form factor of 
 $T^\mu_\mu$.    

The classification of theories in any dimension 
 with rational $\cfp$ is shown to be
very closely
related to the construction  of polylogarithmic ladders in
mathematics\cite{Lewin}, in particular classical analysis and
 number theory.    Various aspects of this are explored
in sections IX-XI.   Given a rational $\cfp$ that can be obtained
from the higher dimensional 
polylogarithmic expressions of section VII, one can 
work back-wards by reconstructing the algebraic gap equation.
The latter gives some information on the interaction parameters.
For integrable theories in 2D, due to the tight constraints of
integrability, one can in many cases entirely reconstruct the 
S-matrix of the theory.  In higher dimensions we do not as yet
have the tools necessary to  complete
the reconstruction of the  fixed point theory. However  we believe
our analysis points to the existence of interesting conformal
theories in higher dimensions and we can at least reconstruct
some of their properties, such as particle spectrum and interaction
parameters.

An interesting example worked out in detail is that of a single
bosonic particle with gap parameter $g=\log 2$.  The polylogarithmic
identities leading to $\cfp = 1/2$ in 2D and $\cfp = 7/8$ in 3D where
known to Euler and Landen.    
Though $\cfp = 1/2$ is the central charge of a  free fermion in 2D, 
the  physical meaning of our bosonic theory and its possible connection
to the Ising model will remain  unanswered.      
The reason it  
has not been studied before  is that previously nobody has been
able to make sense of a {\it bosonic} quantum statistical
mechanics in the TBA framework\cite{Mussbos}.

In section XII it is shown how a purely imaginary chemical potential
can describe fractional statistics particles.  In this construction
it is necessary to work with gap parameters in the complex plane
$z = e^{-g}$.  For free ``anyons'' the gap parameters $z$ are
on the unit circle.    In 2D the free anyons correspond to 
$\beta-\gamma$ systems in a continuous sub-regime of the spin 
$\sigma = \vartheta/2\pi$  of
the particles where $0<\vartheta<2\pi$. 
When the spin is rational, so is $\cfp$.   
Extending this construction to 3D we propose a formula for
the free energy of free anyons.  Here $\cfp$ is only rational
for special values of the statistics parameter such as 
$\vartheta/2\pi = 1/3$.  

In section XIII we reintroduce the chemical potential 
and analyze the phase structure.   This leads to general statements
concerning the existence of fixed points depending on the 
interaction strengths and chemical potential.

\section{Free energy and the stress-energy  tensor}

In this section we introduce the main physical  quantities that 
are the subject of this paper; we mainly collect some well-known
formulas.     From the most general
point of view, we will be dealing with a euclidean functional integral
of fields in $D=d+1$ dimensions where one of the dimensions
is compactified into a circle of circumference $\beta$.  
This object arises in a variety of physical contexts.  
It is the finite temperature $T$ partition function of
a quantum many-body system or quantum field theory in $d$ spacial
dimensions where the extra dimension is the euclidean time and 
$\beta = 1/T$.   It can also represent a partition function
of classical statistical mechanics in $D$ spacial dimensions
with one finite size direction of length $\beta$.  
We will mainly be using the language of quantum statistical mechanics.  
Most formulas will be expressed in terms of $\beta$, but some
formulas will be expressed in terms of the temperature $T$ when
this makes the physical context more transparent.  Some formulas
will be expressed in terms of $D$, others in terms of $d$, depending
on what is more convenient.     

The partition function is
\beq
\label{1.1}
Z (\beta , \mu )  = \Tr ~ e^{-\beta (H-\mu \hat{N} )}
\eeq
where $H$ is the many-body hamiltonian, $\hat{N}$ the particle 
number operator, and $\mu$ the chemical potential.   
The free energy density, $\CF = F/V$,  is  defined as:
\beq
\label{1.2}
\free = - \inv{\beta V} \log Z 
\eeq
where $V$ is the $d$-dimensional spacial volume.   
The energy density $\CE$  per volume and pressure $p$  are defined as 
\beq
\label{1.3}
\CE = - \inv{V} \dd{\log Z}{\beta} = \dd{(\beta \CF)}{\beta}
  , ~~~~~p = \inv{\beta} \dd{\log Z}V = - \CF 
\eeq

The above quantities are related to the stress-energy tensor
of the theory $T_{\mu \nu} (\xvec ) $,  $\mu, \nu = 0,1,..,d$.  
The hamiltonian $H$ and momentum operators $\vec{P}$ are 
given by  
\beq
\label{1.4}
H = \int d\xvec ~ T_{00} (\xvec ) , ~~~~~~P_i = \int d\xvec ~ T_{0i} (\xvec)
\eeq
where $d \xvec \equiv d^d \vec{x}$.  
The energy density and pressure are related to finite temperature
one-point functions of $T_{\mu\nu}$:
\beq
\label{1.5}
\langle T_{00} \rangle_\beta = \CE, ~~~~~ \langle T_{ii} \rangle_\beta 
= p
\eeq
(no sum on $i$), 
where $\langle * \rangle_\beta$ denotes the finite temperature 
correlation function. (The one-point functions are independent of 
$\xvec$ by translation invariance.) 
The trace of the stress-energy tensor,
$T_\mu^\mu  \equiv  \sum_{\mu = 0}^d  T^\mu_\mu$,
can thus be related to the free energy:
\beq
\label{1.6}
\langle T_\mu^\mu \rangle_\beta = ( \beta \d_\beta + d + 1 ) ~ \free
\eeq

In the sequel we will also deal with the entropy density $\CS$ 
and number density $n$:
\beq
\label{ns}
\CS = \beta^2 \dd\free\beta = \beta ( \CE - \free ), ~~~~~n = - \dd\free\mu
\eeq
The first equation is just the first law of the thermodynamics.

\def\cqstat{{C_{\rm qs}}}
\def\cfp{{c_{\rm qs}}}

\section{Quantum statistical  c-functions in D dimensions}

\def\kvec{{\bf k}}

\def\omk{\om{\kvec}}

\def\dk{\frac{d\kvec}{(2\pi)^d}}

\def\smallentropy{{\rm e}}

For a free gas of bosons or fermions, 
the partition function is well-known in the limit of infinite
volume $V$: 
\beq
\label{3.1}
\log Z = \mp V \int \dk ~ \log \( 1 \mp e^{-\beta \omk } \) 
\eeq
where the upper (lower) sign corresponds to bosons (fermions),
 $d\kvec \equiv d^d \vec{k}$,
and $\omk$ is the single particle energy as a function of the
momentum $\kvec \equiv \vec{k}$, which  need not be relativistic.
For a massless relativistic theory with $\omega_\kvec = \sqrt{\kvec^2}$,
the integrals for the basic thermodynamic quantities can be done
analytically.  
  For example in 4D, the energy density is the well-known
black body formula:
\beq
\label{freegas1}
\CE = \cfp \frac{\pi^2}{30} T^4
\eeq
where $\cfp = 1$ for a free boson and $7/8$ for a free fermion. 
One can also relate the entropy $S$ and the number of particles $N = 
n V$:
\beq
\label{freegas2}
S =  \smallentropy _\mp  ~ N
\eeq
where $\smallentropy_-$ ($\smallentropy_+$) is the entropy 
per particle for bosons (fermions):
\beq
\label{freegas3}
\smallentropy_- = \frac{(d+1) \zeta (d+1)}{\zeta (d)} , 
~~~~~~
\smallentropy_+ = \smallentropy_- \( \frac{2^d-1}{2^d - 2} \)  
\eeq
and  $\zeta (z)$ is the Riemann zeta function. 
Finally the ideal gas law reads:
\beq
\label{idealgas}
pV = \frac{\smallentropy_\mp}{d+1} \, NT
\eeq
  
One question that will be addressed in this paper is how interactions
can change the coefficient $\cfp$ in eq. (\ref{freegas1}) 
even at a fixed point. 
This leads us to  introduce a quantum statistical c-function,  $\cqstat (T)$, 
which depends on temperature,  and at a fixed point becomes in 4D for example
$\cfp$ in eq. (\ref{freegas1}). 
This quantity can be defined in
any dimension.     
If the theory is relativistic and  conformally invariant,  the trace of the 
stress tensor vanishes $\langle T_\mu^\mu \rangle_\beta = 0$.  
Equation (\ref{1.6}) then implies that $\free = {\rm const.}/\beta^{d+1}$.  
This leads to a natural definition of a thermal  c-function  based on 
quantum statistical mechanics: 
\beq
\label{1.7}
\free = -  \frac{a_d}{\beta^{d+1}} ~ \cqstat (\beta, \lambda )
\eeq
where $a_d$ is a normalization constant.  The function
$\cqstat$ depends on $\beta$ and the couplings and masses
of the theory, here  denoted generically as $\lambda$.
Because the definition eq. (\ref{1.7}), which is essentially
motivated simply by  dimensional analysis, is so natural, it has
 previously appeared in a 
number of works\cite{Gennes,Neto,Sachdev,Petkou,Appelquist,Petkou2,Gaite}.
(In \cite{Gennes} the finite size, rather than finite temperature, version
was considered.)    
This paper is not so much concerned with proving or disproving
c-theorems, but rather
with developing 
a new framework for computing $\cqstat$.  Nevertheless, 
some remarks on c-theorems
can be found later in this section.

The trace of the stress-energy tensor obeys renormalization
group equations (RG)  since it is a one-point function of
a quantum field theory on a finite size geometry, the length
scale being $\beta$.  Thus:
\beq
\label{RGflow}
\Theta ( e^s \beta , \lambda (l)) = e^{-Ds} \Theta ( \beta, \lambda (l+s) ), 
~~~~~~~
\Theta (\beta , \lambda ) \equiv \langle T_\mu^\mu \rangle_\beta
\eeq
Above, $l$ is the log of the length scale, and the $l$
dependence  of $\lambda$ is determined by the 
RG beta-functions: $d\lambda /dl = \dot{\lambda}  (\lambda)$. 
The above equation shows that $\cqstat(\beta)$
as a function of $\beta$ tracks the RG flow.   Using eq. (\ref{1.6})
one has:
\beq
\label{RGflow2}
\Theta (\beta) = \CE - d p =   {a_d} T^{d+2}   ~ \dd\cqstat{T}
\eeq
Thus at an RG fixed point $\Theta = 0$,  $\cqstat$ is 
a constant independent of $\beta$,  and is an important characteristic of
the fixed point theory.  We will use the lower case $\cfp$ to denote
the $\beta$ independent fixed point value of $\cqstat$.

The entropy and energy densities in terms of $\cqstat$ are:
\beq
\label{csc}
\CS = \frac{a_d}{\beta^d} \( (d+1)  - \beta \d_\beta \) \cqstat 
\eeq
\beq
\label{Ecc}
\CE = \frac{a_d}{\beta^d} \( d  - \beta \d_\beta \) \cqstat 
\eeq
At a fixed point   one  has:
\beq
\label{ce}
\CE =  d \frac{ a_d}{\beta^{d+1}} ~ \cfp , 
~~~~~\CS = (d+1)  \frac{ a_d}{\beta^{d}} ~ \cfp
\eeq
so that $ \CE/\CS = \frac{d}{d+1} T$. 

We  chose $a_d$ so that $\cfp = 1$
for a free, massless relativistic boson.  As we will show below, this
leads to:
\beq
\label{1.8}
a_{D-1}  \equiv  \pi^{-D/2}  \Gamma (D/2) \zeta (D)
\eeq
where $\Gamma$ is the gamma-function and  as before
$\zeta$ the Riemann zeta-function.
When  $D$  is even, $\zeta (D)$ is rational up to powers of $\pi$:
$\zeta (D) = 2^{D-1} \pi^D |B_D|/D! $, where $B_D$ are rational
Bernoulli numbers.  This is not the case for $D$ odd;  in
fact it has only been proven relatively recently that $\zeta (3)$ 
is irrational.  We checked numerically that $\zeta(D)/\pi^D $ for $D$ odd 
doesn't appear to be rational.  
    As we explain below, 
this appears to be related to the fact that there is no conformal
anomaly for $D$ odd.  
For $D=2,3,4$ the expression for the free energy then becomes:
\barray
\nonumber
\free &=& - \cqstat (T) ~ \frac{\pi T^2}{6} ~~~~~~~~~~~~~(D=2) 
\\  
\label{1.9}
 &=& - \cqstat (T) ~ \frac{\zeta (3) T^3 }{2\pi}  ~~~~~~~~~~(D=3)
 \\
\nonumber
 &=& - \cqstat (T)  \frac{\pi^2 T^4 }{90} ~~~~~~~~~~~~~(D=4)
\earray

Since in recent years there has been renewed interest in possible
generalizations of Zamolodchikov's c-theorem\cite{Zctheorem} to
higher dimensions\cite{Cardyc4,Anselmi,Intriligator,Csaki}, 
we now discuss the relation of $\cfp$ to other proposals in
higher dimensions.  Let us first review the success story in 
$D=2$.   In a conformally invariant theory,  $c$ can be defined
by the leading singularity of the operator product expansion of
two stress-energy tensors,  or equivalently the 
central extension in the Virasoro algebra\cite{BPZ}. 
The same central charge $c$ appears as a conformal anomaly in the
conformal field theory:
\beq
\label{top1}
\langle T_\mu^\mu \rangle = - \frac{c}{48 \pi} R
\eeq
where $R$ is the scalar  gravitational curvature.  Using the above relation 
one can show that $c$ governs the finite size effects\cite{Cardy,Affleck}
 of the partition function and thus at a fixed point
$c$ is the same as  $\cfp$.

   For the purpose of studying the RG flow between
fixed points there are two known alternative approaches to defining
a scale dependent $c$ that tracks the flow and equals the Virasoro
central charge at the fixed point.  One is to define $\cqstat (\beta )$ 
as above\cite{ZamoTBA}.   The other is the content of the c-theorem where one
defines a function $c_z (L)$ from the two-point functions of 
$T_{\mu\nu}$ at zero temperature but at finite separation $L$.  
The c-theorem states that $c_z (L)$ monotonically decreases as
a function of the increasing length scale $L$.   There is sometimes the  
misconception that  $\cqstat$ and $c_z$ 
are essentially the same.   
This misconception  has led to the expectation that $\cqstat$ also always 
decreases, even in 2D.  
However these two c's are potentially very different,  as one
is related to a one-point function at finite temperature, the other
a two-point function at zero temperature.  The thermodynamic foundation
for $\cqstat$ actually makes it a much richer physical quantity,
and obviously one that can be defined in any dimension.

Though some discussions of the c-theorem  vaguely invoke   
notions in statistical physics like irreversibility and loss of
information,  in reality $c_z$ has
no statistical mechanical meaning.  The intuitive
understanding is more accurately that $c_z$ is a measure of the massless states
of the theory, and since massive states decouple at low energies,
$c_z$ decreases\cite{Friedan}.
   Ironically, whereas $\cqstat (\beta )$ 
does have a thermodynamic meaning,  no theorem has ever
been proven for it. 
     Indeed, though $\cqstat$ is known to decrease
for theories with UV and IR 
fixed points, it can actually oscillate in physically
sound  theories
with RG limit cycles\cite{LRS,LSelliptic}.

\def\Tuu{T^\mu_\mu}

As eq. (\ref{RGflow2}) obviously 
shows, a  c-theorem for $\cqstat$  follows from
positivity of $\Theta$:  if $\Theta > 0 $, then $\d \cqstat/ \d T  >0$, i.e. 
$\cqstat$ increases with temperature.  Let us refer to the above condition
as $\Theta$-positivity.  In \cite{Appelquist} many
interesting examples were presented that exhibit the
expected property $\cfp^{IR} < \cfp^{UV}$; however examples where 
$\cqstat$ does not decrease monotonically with decreasing temperature
were also discussed.   

  A general manner in which $\Theta$-positivity
can be violated is as follows\cite{Appelquist,LSelliptic}.
  Suppose a quantum field theory
is  described by
a conformally invariant  action $S_{\rm cft}$  perturbed  by
 operators  $\CO^A$:
\beq
\label{S00}
S = S_{\rm cft}  + \sum_A  \int \frac{d^2 x}{2\pi} ~  \lambda_A  \CO^A 
\eeq
where $\lambda_A$ are positive couplings.
  Let $\dot{\lambda}_A$ be the beta function for
$\lambda_A$ and $\Gamma_A$ the scaling dimension
(including anomalous)  of $\CO^A$.  Then the
beta functions  $\dot{\lambda}_A = d\lambda_A / d\log L$, where $L$ is
a length scale,  to lowest order are 
\beq
\label{betass}
\dot{\lambda}_A = (D-\Gamma_A ) \lambda_A + O(\lambda^2) 
\eeq
The main point is that the trace of the stress-energy tensor 
receives quantum corrections, and since it must be zero at a fixed 
point where the beta functions are zero, one must have:
\beq
\label{Tbeta}
T_\mu^\mu = \sum_A \dot{\lambda}_A (\lambda)  \, \CO^A
\eeq
The above formula is well known and is easily 
 verified in 2D to lowest order in conformal perturbation theory\cite{ZamoRG}. 
Consider first a theory that can be formulated as a perturbation of
an ultra-violet fixed point  by relevant operators, which
implies $\Gamma_A < D$.  Then the beta-functions are positive 
to lowest order.  Furthermore, for relevant perturbations, because
of the anomalous dimensions of the couplings, 
 there is often 
no higher order corrections to the beta functions since higher powers
of $\lambda_A$ do not have the right dimension.   So in this
situation, $T_\mu^\mu$ is generally positive and the c-theorem holds.  

The above arguments clearly point to the way in which the c-theorem
can be violated. If the  $\CO^A$ are marginal, $\Gamma_A  = D$, 
and the beta functions start at $O(\lambda^2)$, and there are no
general constraints on the sign of the beta function. 
This was suggested as  the origin of the violation of the c-theorem
for the 2D models with limit-cycles in the RG\cite{LRS,LSelliptic},
though the issue is not entirely resolved.

What  constraints  on the flow of $\cqstat$ follow from the most basic
thermodynamic principles?  Note that the entropy itself, and not
its change,  already depends on the flow of $\cqstat$, 
eq. (\ref{csc}).    One constraint is that the entropy density should
be positive.  Since the pressure is proportional to $\cqstat$, 
there are two cases depending on the sign of $\cqstat$.
Using eq. (\ref{csc}) and (\ref{Ecc}) 
(in terms of temperature $T$):
\bigskip
\barray
\label{boundplus}
{\rm {\bf{positive~ pressure:}}} ~~~~~~&\CS& > 0 ~~~~~~\Longrightarrow ~~~~~~ 
\dd{\log \cqstat}{\log T} > -d-1
\\
\nonumber
~~~~~~&\CE& > 0 ~~~~~~\Longrightarrow ~~~~~~ 
\dd{\log \cqstat}{\log T} > - d
\earray
\bigskip
\barray
\label{boundminus}
{\rm {\bf{negative ~ pressure:}}} ~~~~~~&\CS& > 0 ~~~~~~\Longrightarrow ~~~~~~ 
\dd{\log \cqstat}{\log T} < -d-1
\\
\nonumber
~~~~~~&\CE& > 0 ~~~~~~\Longrightarrow ~~~~~~ 
\dd{\log \cqstat}{\log T} < - d
\earray
Note that in the positive pressure case, a positive entropy
could have a positive or negative energy density.  On the other hand
in the negative pressure case, a positive entropy necessarily
implies a positive energy density.

Let us now turn to the conformal anomalies in higher
dimensions.    In $D$ dimensions $T_\mu^\mu$ 
has scaling dimension $D$ and the curvature $R$  has dimension $2$.  
This already indicates that there is no conformal anomaly for 
$D$ odd since no power of the curvature has dimension $D$.
As mentioned above, this appears to be reflected in the formulas
eq. (\ref{1.9}) by the irrationality of $\zeta (D)$ up to geometrical powers
of $\pi$ for $D$ odd.   Again the  quantity $\cqstat$ presents itself 
as the most natural 
analog  to $c$  for all even and odd dimensions. 

For $D=4$ the conformal anomaly reads\cite{Duff,Anselmi} 
\beq
\label{anomaly}
\langle T_\mu^\mu \rangle =  \inv{1920 \pi^2} 
\( \hat{c} \,   W^2 -  a  \tilde{R}^2 \) 
\eeq
where $W$ is the Weyl tensor and $\tilde{R}$ the dual curvature. 
(We have rescaled $\hat{c},a$ so that $\hat{c}=1$ for a free boson.) 
Cardy conjectured an $a$-theorem for the anomaly $a$, 
and recently there have been many interesting examples of 
supersymmetric gauge theory supporting the conjecture\cite{Anselmi,
Intriligator,Csaki}.  Many of these examples rely on Seiberg 
duality. 
It should be emphasized that the above equation is valid only
at a fixed point, and thus does not provide a definition of
scale-dependent $\hat{c},a$ that interpolate between fixed points.  
Since  $\hat{c}$ and $a$ are related to one-point functions of
$T_\mu^\mu$, they are analogous to $\cfp$ rather than $c_z$.  
However one can easily show that no linear combination of $\hat{c},a$
equals $\cfp$.  Consider a free theory consisting of 
$N_0$ scalars, $N_{1/2}$ spin $1/2$ Majorana fermions, and
$N_1$ spin $1$ gauge bosons.   The anomalies are known\cite{Duff,Anselmi}:
\barray
\nonumber
\hat{c}  &=& N_0 + 3N_{1/2} + 12 N_1 
\\
\label{caanom}
a &=& (2N_0 + 11 N_{1/2} + 124 N_1 )/6 
\earray
On the other hand $\cfp$ only depends on the statistics
of the particles and their number of degrees of freedom, 
which are 2 helicities for the fermions and 2 polarizations
for the gauge bosons.  Using $\cfp = 1$ for a boson
and $\cfp = 7/8$ for a fermion:
\beq
\label{cstatpart}
\cfp = N_0 + \frac{7}{4} N_{1/2} + 2 N_1
\eeq
Noting that $3(\hat{c}-a)/2 = N_0 + 7 N_{1/2} / 4 - 13 N_1 $,
one concludes that in a free theory of only scalars and spin $1/2$ 
particles, $\cfp = 3(\hat{c}-a)/2$, however for higher spin particles
there is no linear combination of $\hat{c}$ and $a$ that equals 
$\cfp$.

Since the equation (\ref{anomaly}) involves the curvature,
the anomalies $\hat{c},a$ can determine some finite size effects.
However  unlike the $D=2$ situation,  $\hat{c},a$ apparently
do not govern the finite size effects having to do with finite
temperature. 
  Indeed,
lattice simulations of finite temperature pure QCD observe
the behavior eq. (\ref{freegas1})  at high temperature
with $\cfp = 16$ which is consistent with an $SU(3)$ gauge
theory with $8$ gauge bosons\cite{Karsch}. The fact that 
$\cfp$ is here given by the free field values is an
indication of asymptotic freedom of QCD\cite{Wilczek}.

\section{Quantum Statistical C-function in Cosmology}

The  free field fixed point values of $\cqstat$
are used extensively in
studies of the thermodynamics of the hot early universe.
For instance the energy density is taken to be given by the black body
formula (\ref{freegas1})  
for a radiation dominated universe. 
This is clearly an approximation in a number of respects. 
First of all, even at a fixed point, the interactions can
change the value of $\cfp$, unless the theory
is asymptotically free.    This will be the subject of
section VII 
of this paper.  Secondly, the universe is probably
not at a fixed point.  Our comments on cosmology will be
confined to this section, where we primarily  phrase some
cosmological properties in terms of $\cqstat$ and use this
to place bounds dark energy that is not a cosmological constant. 
In other words, we place bounds on dark energy that is hypothetically
made up
of particles.

\def\w{{ w}}
\def\Rdotdot{\ddot{R}}

There is growing observational evidence for an acceleration of the expansion
of the universe.  The cause of this phenomenon is commonly  referred to
as dark energy.   (See for instance, \cite{Turner,Ratra,Carrol}.)   
Dark energy  is usually characterized by
the equation of state  parameter $\w$:   
\beq
\label{dark1}
\w \equiv  \frac{p}{\CE}  
\eeq
One of the Friedmann  equations expresses the acceleration in terms
of $\CE, p$:
\beq
\label{Freedman}
\frac{\Rdotdot}{R} = - \frac{4\pi G}{3} (\CE + d p ) =  - \frac{4\pi G}{3}
\( \inv{\w} + d \) p 
\eeq
where throughout this section 
$d=3$ and $R$ is the cosmic scale factor.  Accelerated expansion,
i.e. dark energy, 
corresponds to $\Rdotdot/{R} > 0$.  If the pressure is positive, 
then  dark energy requires $-1/d < \w < 0$ which corresponds to a
negative energy density.   If on the other hand the pressure is 
negative, dark energy requires $\w < -1/d$ with positive energy 
density, or $\w > 0 $ but with negative energy density.

The parameter $\w$ can  be expressed in terms of $\cqstat$:
\beq
\label{dark2}
\inv{\w} =  d + \dd{ \log \cqstat}{\log T}
\eeq
At a fixed point, $\w=1/d$ regardless of the value of $\cfp$.
However away from a fixed point, it can be different. 
Already one sees that if dark energy is matter, it is currently
not at an RG fixed point.  
The thermodynamic bounds coming from the
requirement of positive entropy,  eqns. (\ref{boundplus},\ref{boundminus}), 
place bounds on $\w$:  
\barray
\nonumber
{\bf {\rm positive ~ pressure:}} &~~~~~~&\CS > 0 \Longrightarrow 
~~~~~ \w < -1 ~~~{\rm or } ~~~ \w >0
\\
\label{darkandlovely}
{\bf {\rm negative ~ pressure:}} &~~~~~~&\CS > 0 \Longrightarrow 
~~~~~ -1 < \w < 0
\earray
Thus one sees that positive pressure dark energy, which
required $-1/d < \w < 0$,  is inconsistent
with positive entropy.  More importantly, taking into account 
the entropy bounds, negative pressure dark energy necessarily
corresponds to $-1 < \w < -1/d$.
  
The standard candidate for dark energy is a cosmological constant
$\Lambda$, where the stress tensor $T^{\mu\nu} \propto  \Lambda g^{\mu\nu}$.
In Minkowski space, this can correspond to the stress-energy tensor
of vacuum energy.    From the Lorentz structure of the flat space 
metric $\eta^{\mu\nu} = {\rm diag} (1, -1,-1, -1)$,
 this
necessarily implies $p=-\CE$ and $\w=-1$.  Interestingly, 
the lower limit $\w = -1$, which  we obtained from thermodynamic
principles, precisely mimics a cosmological constant. 

If dark energy is not a cosmological constant,  then in terms of
$\cqstat$ the most likely scenario for dark energy is that 
the universe is  currently at a temperature where $\cqstat$ is negative
with $\d \cqstat / dT >0$, i.e. $\cqstat$ is decreasing as the
temperature continues to decrease.  The fate of the universe then
depends on whether there is an RG fixed point at zero temperature.
If there is a fixed point, then as explained above,  $\w$ will eventually
become positive and equal to $1/3$.  

As we will see, obtaining a negative $\cqstat$ is not so straightforward
in our formalism.   One can study this by examining fixed point values
$\cfp$ analytically and see what is required for $\cfp$ to be negative,
since $\cqstat$ is bound to be negative in the vicinity of the fixed
point.   As we will show in section XIII,  this seems to require 
an imaginary part to the chemical potential.  One interpretation 
of an imaginary chemical potential in 2D and 3D  will be given in section XII, 
i.e. as corresponding to fractional statistics particles. 
Analogous Hopf terms in 4D would require $Im(\mu) = i\pi T$ 
since there are only bosons and fermions in 4D;  one example is  
skyrmions\cite{Witten}. 
But there
may be other interpretations to an imaginary part,  for example 
a relaxation rate.

\def\omk{\om{\kvec}}

\def\dk{\frac{d\kvec}{(2\pi)^d}}

\section{Quasi-particle re-summation}

This section is rather technical, so let us begin by
drawing attention to the main results:  eq. (\ref{Z}) is
the desired form of the partition function,  where $\vep$
satisfies the integral gap equation (\ref{pseudo}).   
The kernel appearing in this integral equation is 
given by the form factor eq. (\ref{kernelT}).

Interacting field theories,  even at finite temperature, normally
have a particle description.  In many complex condensed matter
systems one is accustomed to dealing with quasi-particle excitations 
of definite statistics,
for example in Landau's theory of a  Fermi liquid for metals.   The 
free particle form of $\log Z$, eq. (\ref{3.1})
 is appealing since it incorporates
the bosonic or fermionic statistics of the particles.
The eq. (\ref{3.1}) can of course be obtained from a
functional integral in finite size which involves the 
Matsubara frequencies, but only with some effort;  the
simplicity of the final answer suggests there is a simpler, particle
description.  
Based on these observations, we  formulate a quasi-particle
description of $Z$ with the following properties.  We suppose
that the free particle form eq. (\ref{3.1}) continues to hold
but with $\om{\kvec}$ replaced by a quasi-particle energy
$\vep  (\kvec )$.  All the effects of the interactions, and additional
finite temperature effects are all contained in $\vep$.

There are no general arguments showing that finite temperature field
theory can be organized in this fashion.  The formulation  should be viewed 
as a potentially useful approximation that can capture some physical
effects in a transparent way in comparison with other approaches.  
However for 2D integrable theories this formulation is exact
and known  as the thermodynamic Bethe ansatz, and this certainly
provided some motivation and insights.

\def\Gkern{{\bf G}}
\def\Kkern{{\bf K}}
\def\2pi{(2\pi)^d} 
\def\entropy{\$} 

We first describe a particle description of the partition function.
To proceed, introduce a density of particles $\rho (\kvec )$ so that 
$\rho (\kvec ) d\kvec$ represents the number of particles
with momentum  between $\kvec$ and $\kvec + d\kvec$.   Similarly
introduce the level density $\rho_l (\kvec )$ which counts the
available quantum states.   In a free theory $\rho_l$  just counts
the states for particles in a box of volume $V$ and 
$\rho_l = V/(2\pi)^d$.  In an interacting theory the densities 
$\rho$ and $\rho_l$ are clearly not independent:  when a new particle is 
added it affects the available levels because of the interactions.  
For integrable theories in 2D  the quantization condition on
allowed momenta leads to a relation between $\rho$ and $\rho_l$ 
involving the 2-particle factorized S-matrix\cite{Yang,ZamoTBA}.
(See section XB.)  
It is not clear that this kind of  quantization condition can
actually be formulated in 
higher dimensions.  So let us for now simply assume that there
is a linear relation between the densities that depends on 
the momenta.  This can generally be written in terms of 
a temperature independent Fredholm integral operator (kernel) $\Gkern$:
\beq
\label{3.2}
\rho_l (\kvec ) = \frac{V}{(2\pi)^d } +  \int \frac{d\kvec'}{(2\pi)^d} 
~ \rho (\kvec' ) G(\kvec', \kvec ) \inv{\om{\kvec}} 
\eeq

\def\fill{{f}}
\def\dkline{\underline{\underline{d\kvec}}}
\def\dklinep{\underline{\underline{d\kvec'}}}
\def\dklineind#1{\underline{\underline{d#1}}}

The kernel $\Gkern$ will be determined below. 
However as we now show, 
  {\it any} choice of $\Gkern$ leads to the desired form
of the partition function.   Retaining the chemical potential, one has
\beq
\label{3.3}
\log Z = - \beta \int d\kvec  ~ ( \om{\kvec} - \mu ) \rho (\kvec ) + \entropy
\eeq
where $\entropy$ is the entropy.  The entropy depends on the 
statistics of the particles: 
\beq
\label{3.4}
\entropy = \pm \int d\kvec \( (\rho_l \pm \rho )\log (\rho_l \pm \rho ) 
- \rho_l \log \rho_l \mp \rho \log \rho \) 
\eeq
where the upper/lower sign is for boson/fermion statistics.
(In the subsequent formulas the boson case will always be the 
upper sign.) 
We now look for a saddle point approximation to $Z$.  Varying
with respect to both densities:
\beq
\label{3.5}
\delta \log Z = \( -\beta ( \omega - \mu )  + \log(\rho_l/\rho \pm 1)\) 
\delta\rho 
\pm \log(1\pm \rho/\rho_l ) \delta \rho_l 
\eeq
The saddle point is subject to the constraint 
eq. (\ref{3.2}) relating $\rho, \rho_l$.    The saddle point
equation $\delta \log Z = 0$ then has a simple form.  Define
the filling fractions
\beq
\label{3.6}
\fill \equiv \frac{\rho}{\rho_l} 
\eeq
and parameterize them as follows:
\beq
\label{3.7}
\fill_\mp = \inv{ e^{\beta \vep} \mp 1 } 
\eeq
Then the saddle point equation becomes a non-linear integral
equation involving $\Gkern$:
\beq
\label{pseudo}
\vep (\kvec ) = ( \om{\kvec} - \mu )  \pm \inv{\beta} 
\int \dklinep ~ G(\kvec , \kvec') ~ \log \( 1\mp e^{-\beta \vep (\kvec ' )} \) 
\eeq
where we have defined:
\beq
\label{3.8}
\dkline \equiv \frac{d\kvec}{\2pi \om{\kvec} } 
\eeq
which is Lorentz invariant when the theory is relativistic.  
Plugging this back into $\log Z$ one finds that it has
the desired form:
\beq
\label{Z}
\log Z = \mp V \int \frac{d\kvec}{\2pi} ~ \log \( 1\mp e^{-\beta \vep (\kvec)
 } \) 
\eeq
We will refer to the equation (\ref{pseudo}) as the
{\it integral gap equation}.

The fundamental thermodynamic quantities, obtained by differentiation,
have  simple expressions:
\barray
\label{EN}
\CE &=& \int \frac{d\kvec}{\2pi} ~ f_\mp (\kvec ) ~ (1+ \beta \d_\beta) \vep
\\
\label{EN2}
n &=& - \int \frac{d\kvec}{\2pi} ~ f_\mp (\kvec ) ~ \dd\vep\mu 
\earray
Also, using $\grad\kvec \cdot \kvec = d$, one has for the pressure
\beq
\label{press}
 p =   \inv{d} \int \frac{d\kvec}{\2pi} ~ f_\mp (\kvec ) ~
\kvec \cdot \grad\kvec \vep (\kvec ) 
\eeq

Using the integral equation satisfied by $\vep$, one can obtain 
an integral series representation for the above quantities.  
Introduce the shorthand notation
\beq
\label{3.11}
(\Gkern  * \phi  ) (\kvec ) \equiv \int \dklinep ~ G(\kvec , \kvec' ) 
\phi (\kvec' ) 
\eeq
for an arbitrary function $\phi (\kvec )$.
(Where  there is no chance for confusion we will sometimes drop the 
$*$.)
Then using the integral equation (\ref{pseudo}) one can show:
\beq
\label{3.12}
(1-\Kkern)(1+\beta \d_\beta ) \vep = \omega
\eeq
where $\Kkern$ is the kernel
\beq
\label{Kkernel}
K( \kvec , \kvec' ) = G(\kvec , \kvec' ) \fill_\mp  (\kvec' ) 
\eeq
Similarly
\beq
\label{numK}
(1-\Kkern) \d_\mu \vep  = -1
\eeq
Using now $(1-\Kkern)^{-1} = \sum_{m=0}^\infty \Kkern^m $ one obtains
\beq
\label{esum}
\CE
 = \sum_{m=0}^\infty \int \frac{d\kvec}{\2pi} ~ \fill_\mp (\kvec) ~
    \Kkern^m * \omega
\eeq
\smallskip
\beq
\label{seriesn}
n =  \sum_{m=0}^\infty \int \frac{d\kvec}{\2pi} ~ \fill_\mp (\kvec) 
~\Kkern^m * 1
\eeq

So far, our equations are quite general and 
 valid whether the underlying theory
is Lorentz invariant or not.
For the remainder of this article, 
we assume the theory is Lorentz invariant with:
\beq
\label{omeg}
\omega_\kvec = \sqrt{\kvec^2 + m^2}  
\eeq
We can now derive an expression we will need for the pressure. 
Introduce the 4-vectors $k_\mu = (\omega_\kvec , \kvec)$ with 
$k^\mu k_\mu = m^2$.  The kernel $\Gkern$ can only be a function of
the Mandelstam variable:
\beq
\label{Mand}
s (\kvec, \kvec' ) \equiv (k_\mu + k'_\mu )^2 = 2 (m^2 + k^\mu k'_\mu )
\eeq
(The other possible variable $t = (k_\mu - k'_\mu )^2 = 4m^2 - s$ is not independent.)
We will need
\beq
\label{grads}
\grad{\kvec} s = \frac{2}{\omega_\kvec} (\omega_{\kvec'} \kvec - \omega_\kvec \kvec' ), 
~~~~~
\grad{\kvec'} s = \frac{2}{\omega_{\kvec'}} (\omega_{\kvec} \kvec'  - \omega_{\kvec'} \kvec )
\eeq
The above  implies for the kernel $\Gkern$:
\beq
\label{4.5}
\om{\kvec} \grad{\kvec} G(\kvec , \kvec' ) = - \om{\kvec'} 
\grad{\kvec'} G(\kvec , \kvec') 
\eeq
Using the integral equation for $\vep$ one can now show
\beq
\label{kinte}
(1-\tilde{\Kkern}) ~ \grad{\kvec} \vep = \grad{\kvec} \omega 
\eeq
where $\tilde{K} (\kvec, \kvec' )
= {\omega_\kvec}^{-1}  K( \kvec, \kvec') \omega_{\kvec'}$. 
This leads to 
\beq
\label{psum}
 p = 
  \inv{d} \sum_{m=0}^\infty \int \frac{d\kvec}{\2pi} ~ \fill_\mp (\kvec) ~
   \kvec \cdot \tilde{\Kkern}^m * \grad{} \omega  
\eeq

\def\Tuu{T^\mu_\mu}

What remains is to determine the kernel $\Gkern$. 
  Consider
the one-point function
$\langle \Tuu \rangle_\beta  $.  From first principles one
has ($\mu = 0$):
\beq
\label{3.9}
\langle \Tuu \rangle_\beta =  \inv{Z} \Tr \( \Tuu  e^{-\beta H} \) 
\eeq
The left hand side can be computed from the free energy eq. (\ref{1.6})
 whereas
the right hand side can be computed by inserting a multi-particle
resolution of the identity into the trace and expressing the 
result in terms of the form factors of $\Tuu$.  In this way
we will relate the kernel to a certain {\it zero temperature} form
factor.   This computation was carried out for 2D integrable theories
in \cite{LecMuss}, and we will subsequently follow this reference
closely.

Let us first determine  $\langle \Tuu \rangle$ from the 
free energy.  Since  $\langle \Tuu \rangle_\beta = \CE + (D-1) \, p$,
using eqs. (\ref{esum},\ref{psum})  one obtains:
\beq
\label{3.13}
\langle \Tuu \rangle_\beta 
 = \sum_{m=0}^\infty \int \frac{d\kvec}{\2pi} ~ \fill_\mp (\kvec) ~
 \(  \Kkern^m * \omega - \kvec \cdot \tilde{\Kkern}^m * \grad{} \omega \)  
\eeq

Next we describe a form-factor computation of $\langle \Tuu  \rangle_\beta$. 
Introduce particle states $|\kvec \rangle$ normalized as follows:
\beq
\label{3.14}
\langle \kvec' | \kvec \rangle =
 \2pi \om{\kvec} ~ \delta^{(d)}(\kvec - \kvec')
\eeq
Inserting a multi-particle resolution of the identity in eq. (\ref{3.9}) 
one finds
\beq
\label{3.15} 
\langle \Tuu  \rangle_\beta = \inv{Z} \sum_{n=0}^{\infty} 
\inv{n!} \int \[ \prod_{i=1}^n  \dklineind{\kvec_i}
 e^{-\beta \om{\kvec_i }} \] 
\langle \kvec_n \cdots \kvec_1 | \Tuu  | \kvec_1 \cdots \kvec_n \rangle
\eeq
The form factors $\langle \kvec_n \cdots| \Tuu |\kvec_1 \cdots \rangle$ 
are the zero temperature form factors.   These form factors have
two kinds of contributions:  disconnected contributions proportional
to $\delta$ functions and the connected part which involves no
$\delta$ functions.  As explained in \cite{LecSal,LecMuss}, the
effect of the disconnected parts is two-fold:  some cancel the
overall $1/Z$ and others convert the $e^{-\beta \omega}$ into 
the filling fractions $\fill_\mp $.  Because the partition function 
involves the quasi-particle energy $\vep$,  the filling fractions $\fill$
are also the quasi-particle energy ones given in eq. (\ref{3.11}).  The final
result is then:
\beq
\label{3.16}
\langle \Tuu  \rangle_\beta = \sum_{n=0}^\infty \inv{n!} 
\int  \[ \prod_{i=1}^n  \dklineind{\kvec_i}  ~ \fill_\mp (\kvec_i )  \] 
\langle \kvec_n \cdots \kvec_1 | \Tuu  | \kvec_1 \cdots \kvec_n \rangle_{\rm
conn} 
\eeq

Finally comparing eq. (\ref{3.16}) with the n-th term in the 
sum eq. (\ref{3.13})  one  finds
\beq
\label{3.17}
\langle \kvec_n \cdots \kvec_1 | \Tuu  | \kvec_1 \cdots \kvec_n \rangle_{\rm
conn} = 
(\om{\kvec_1 } \om{\kvec_n} - \kvec_1 \cdot \kvec_n )
 G(\kvec_n, \kvec_{n-1}) \cdots 
G(\kvec_3 , \kvec_2 ) G(\kvec_2 , \kvec_1 ) + {\rm perm. } 
\eeq
The above factorization is exact for the 2D integrable theories
as a consequence of factorizability of the S-matrix; 
for non-integrable theories it is likely to be an approximation. 
The above equation implies: 
\beq
\label{3.18}
\langle \kvec | \Tuu  | \kvec \rangle_{\rm conn} = m^2   
\eeq
which is a normalization condition. 
The kernel is then  identified with a two-particle to two-particle form 
factor:
\beq
\label{kernelT}
G(\kvec , \kvec' ) = \inv{2 k^\nu k'_\nu  }
~ \langle \kvec , \kvec' | \Tuu  | \kvec' , \kvec \rangle_{\rm conn} 
\eeq
where $2 k^\mu k'_\mu = s - 2m^2$.  

In the sequel we will need a scaling equation obeyed by
the kernel.  $\Gkern$ depends on the momenta $\kvec$ 
and couplings and masses denoted as $\lambda$ with scaling
dimension $d_\lambda$.  Using only the fact that $\Gkern$ has
scaling dimension $(1-d)$ one has:
\beq
\label{Gscaling}
G( e^l \kvec , e^l \kvec' ; e^{d_\lambda l} \lambda  ) = 
e^{(1-d)l}~  G( \kvec, \kvec' ; \lambda )
\eeq

It is straightforward to extend the above formulas to many particles. 
Let there be $N$ particle species, each with statistical parameter
$s_a = \pm 1$, where $s_a = 1$ corresponds to a boson, and 
$m_a$ their masses, $a=1,..,N$.  The equations become:
\beq
\label{many1}
\log Z = - V \sum_a  \int \frac{d\kvec}{(2\pi)^d} ~ 
s_a \log \( 1- s_a e^{-\beta \vep_a} \) 
\eeq
\beq
\label{many2}
\vep_a = ( \omega_{\kvec, a} - \mu_a )  + \inv{\beta} \sum_b  
\Gkern_{ab} * s_b  \log \( 1- s_b e^{-\beta \vep_b} \)
\eeq
where $\omega_{\kvec, a} $ depends on $m_a$.  
The kernels are
\beq
\label{many3}
G_{ab} (\kvec, \kvec') = 
\inv{2 k^\nu_a k'_{\nu , b} }
\langle \kvec, a; \kvec' , b | \Tuu | \kvec' , b; \kvec, a \rangle 
\eeq

In summary, the quasi-particle approach we have developed
is  implemented as follows.  One first computes the kernel $\Gkern$
from the zero temperature form factor eq. (\ref{kernelT}) to whatever
order in perturbation theory is feasible.    
One can then  solve  for the quasi-particle energy $\vep$ 
by numerically solving the integral equation (\ref{pseudo}).  
Finally the partition function is given by the simple eq. (\ref{Z}). 

\section{Comparison with perturbation theory in the Matsubara formalism.}

The standard methodology of finite temperature field theory 
involves perturbation theory where one must sum over Matsubara frequencies. 
Since this looks completely  different from the approach of the last section,
it is important to check that both approaches agree at least to  lowest order. 
In this section we perform this check for the $\phi^4$ scalar
field theory  in any dimension.
The lowest order contribution to the free energy, eq. (\ref{finalF1}) below,
is well-known\cite{dellac,kapusta}.  In the latter
formalism, this lowest order correction is a 2-loop vacuum Feynman 
diagram. 
As we will see,
to obtain this  standard Matsubara result in our formalism  is somewhat
subtle.   

The interacting scalar theory is defined by the lagrangian  
\beq
\label{phi4}
\CL = \inv{2} (\d_\mu \phi )^2 -\frac{m^2}{2} \phi^2 - 
\frac{\lambda}{4!} \phi^4 
\eeq
The usual stress-energy tensor is
\beq
\label{Tstan}
T_{\mu\nu} = \d_\mu \phi \d_\nu \phi - \eta_{\mu\nu} \CL
\eeq
The problem with the usual stress-energy tensor, is that even
in the conformal limit where
 the theory is free and massless ($\lambda, m = 0$), 
the above stress-energy tensor is not traceless.  It is known in this limit
how the stress-tensor can be improved so that it is still conserved but now
traceless\cite{Callan}:
\beq
\label{Tnew}
T^{\rm new}_{\mu\nu} = T_{\mu\nu} + \frac{(d-1)}{2d} 
\( \eta_{\mu\nu} \d^\alpha (\phi \d_\alpha \phi ) - \d_\mu (\phi \d_\nu \phi )
\) 
\eeq

\def\lambdadot{\dot{\lambda}}

\def\omegak{\omega_\kvec}
\def\omegakp{\omega_{\kvec'}}
\def\gradk{\grad{\kvec}}
\def\gradkp{\grad{\kvec'}}

In the interacting case,  
the further improvement eq. (\ref{Tbeta}) is necessary:
\beq
\label{traceT}
T_\mu^\mu =  \frac{\lambdadot}{4!} \phi^4 
\eeq
where $\lambdadot$ 
is the beta function for flow toward the infra-red.  ($\lambdadot$ differs by a sign
from the usual convention of flow toward the ultra-violet in high energy physics.)
Since the field $\phi$ has mass dimension $(D-2)/2$ in $D$ dimensions
and the operator $\phi^4$ has dimension $2(D-2)$,  
\beq
\label{betaf}
\lambdadot = (4-D)\lambda + O(\lambda^2 )
\eeq

We can now 
compute the form factor that determines the kernel.  Expand the field 
in the usual way:
\beq
\label{field1}
\phi (\xvec ) = \int 
\frac{ d\kvec }{(2\pi)^d \sqrt{2\omegak}} 
\( a_\kvec e^{i \kvec \cdot \xvec} + 
a_\kvec^\dagger  e^{-i \kvec \cdot \xvec}
\)
\eeq
where 
\beq
\label{comm}
[ a_{\kvec'} , a_\kvec^\dagger ] = (2\pi )^d ~ \delta^{(d)} (\kvec - \kvec')
\eeq
The states $|\kvec \rangle$ with the normalization in section IX are then:
\beq
\label{states}
|\kvec \rangle = \sqrt{\omegak} ~ a_\kvec^\dagger |0\rangle 
\eeq
Expanding the fields $\phi$ in eq. (\ref{traceT}) and evaluating
 the matrix element
one finds:
\beq
\label{Tlam}
\langle \kvec, \kvec' | \Tuu | \kvec', \kvec \rangle 
= \frac{\lambdadot}{4}
\eeq
Thus the kernel $\Gkern$ to lowest order is 
\beq
\label{Gker4}
G(\kvec , \kvec' ) = \frac{\lambdadot}{4(s-2m^2 )}
\eeq
where $s(\kvec, \kvec' )$ is the Mandelstam variable eq. (\ref{Mand}).

The integral gap equation implies that the lowest order correction to 
$\vep $ is:
\beq
\label{epapprox}
\vep \approx \omegak + \inv{\beta} G * \log (1-e^{-\beta \omega} ) 
\eeq
Inserting this into the free energy, one finds
\beq
\label{Fapprox}
\CF = \CF_0 + \CF_1 + O(\lambda^2) 
\eeq
where 
\beq
\label{F1}
\CF_1 = \inv{\beta}
\int  \dklineind{\kvec} \int  \dklineind{\kvec'} ~ f_0 (\kvec )
\,  \omegak \,  G(\kvec , \kvec' ) \,  \log (1- e^{-\beta \omegakp} )
\eeq
Above, the filling fraction is the free-field bosonic one: 
\beq
\label{f0}
f_0 (\kvec ) = \inv{ e^{\beta \omegak} - 1 } 
\eeq

\def\omegak{\omega_\kvec}
\def\omegakp{\omega_{\kvec'}}
\def\gradk{\grad{\kvec}}
\def\gradkp{\grad{\kvec'}}

Next we express the kernel as a derivative.  For simplicity we only
 consider
 the massless case.   One can easily show:   
\beq
\label{comp1}
\frac{\omegakp}{\omegak} ~ 
\kvec \cdot \grad{\kvec'} \( \inv{s} \) = \inv{s}
, ~~~~~
- \kvec' \cdot \gradkp \( \inv{s} \) =\inv{s}
\eeq
Using these two equations in turn, and integrating by parts, one has:
\barray
\label{subtle}
\CF_1 &=& - \frac{\lambdadot}{4} \int  \dklineind{\kvec} \int 
 \dklineind{\kvec'} ~
f_0 (\kvec ) f_0 (\kvec' ) ~ \frac{\kvec\cdot \kvec' }{s} 
\\
\nonumber
&=& (D-2) \CF_1 + 
\frac{\lambdadot}{4} \int  \dklineind{\kvec} \int  \dklineind{\kvec'}
~ f_0 (\kvec ) f_0 (\kvec' ) ~ \frac{\omegak \omegakp  }{s} 
\earray
The extra $(D-2)\CF_1 $ comes from 
$\gradk \cdot (\kvec / k) = (D-2)/ k $.   
Adding the two above equations one obtains
\beq
\label{finalF1}
\CF_1 = \frac{\lambda}{8} \( \int \dklineind{\kvec} f_0 (\kvec ) \)^2 
\eeq
The above equation is the well-known lowest order correction  to the 
free energy\cite{dellac,kapusta}.  Note that in $4D$ one needs
to keep the leading term $(4-D) \lambda$  in the beta function (\ref{betaf})
through to the last step in order to obtain eq. (\ref{finalF1}).

\section{Thermal gaps and $\cfp$  for interacting fixed points}

\def\zgap{g}

In this section we  analytically
study the integral gap equation for relativistic theories 
 near an  RG fixed point.   In this way
we will obtain  expressions for $\cfp$ for an
interacting fixed point.  Throughout this section the chemical 
potential $\mu = 0$.  For simplicity we first treat  a theory with
a single particle.   

In the approach developed in section V, $\cqstat$ is given by the formula:
\beq
\label{4.2}
\cqstat (\beta ) = \mp \frac{\beta^d}{a_d} \int
\frac{d\kvec}{ (2\pi )^d }~ 
\log \( 1\mp e^{-\beta \vep} \) 
\eeq
As we argued in section III, at a fixed point $\cqstat$ is a
constant independent of $\beta$.   At a fixed point the theory
is expected to be massless.  This can occur in the ultra-violet (UV), i.e.
at high temperature 
or  high energy  compared to the mass 
where $\om{\kvec } \approx \sqrt{\kvec^2} \equiv k$.
We will focus on such  UV fixed points, but some of our analysis can
also apply to infra-red fixed points at low temperature 
  since
such a theory is expected to be massless from the beginning for all $\kvec$
and again $\om{\kvec } =  k$.
The fixed point properties are then determined by the  solution to
the massless integral gap  equation:
\beq
\label{4.1}
\vep  = k ~ \pm~ \inv{\beta} G * \log \( 1\mp e^{-\beta \vep} \)  
\eeq
At a fixed point, $\Gkern$ should depend only on dimensionless
parameters $\lambda$ with $d_\lambda = 0$, 
 so that eq. (\ref{Gscaling}) implies
\beq
\label{Gfp}
 G(e^{l} \kvec , e^{l} \kvec' ) 
= e^{(1-d)l} G(\kvec , \kvec' )
\eeq
Then from eq. (\ref{4.1}) one can show 
\beq
\vep( e^l \kvec ,  \beta ) = e^{l} \vep ( k , e^l \beta )
~~~~~~~({\rm at~a~fixed~point})
\eeq  
The fixed point value of 
$\cqstat$  is then:
\beq
\label{4.2.0}
\cqstat (\beta ) = \mp \frac{\beta^d}{a_d} \int \dkline ~  k ~ 
\log \( 1\mp e^{-\beta \vep} \) 
\eeq
where now $\vep$ satisfies the massless integral  gap equation (\ref{4.1}). 
By rescaling $\kvec$ in the above equation one shows  that
$\cqstat (e^l \beta ) = \cqstat (\beta)$.  In summary,
if $\vep$ satisfies the massless integral gap equation (\ref{4.1}),
and $\Gkern$ the scaling equation (\ref{Gfp}) then 
$\cqstat $ in eq. (\ref{4.2}) is a constant $\cfp$ independent of
$\beta$.  

An important feature of the massless  integral gap  equation (\ref{4.1})
is that $\vep$ generally develops a thermal gap $\Delta \propto T$  
at high temperatures.  When $k$ is small compared to the temperature $T$,
the $k$-term on the RHS of (\ref{4.1}) can be dropped 
and $\beta \vep$ is approximately constant.  Let us define
the dimensionless thermal gap parameter 
 $\zgap$ so that:
\beq
\label{delta}
\vep  \approx g T \equiv \Delta , ~~~~~k \ll T, ~~~~~T ~ {\rm large}
\eeq
   For $k$ much higher than $T$ then $\vep \approx k$.  
 These features are shown in figure 1.   In the limit of $T\to \infty$,
$ \vep = \Delta $ for  all $k$,   and  $\zgap$ is a solution to
the  {\it algebraic gap equation}:
\beq
\label{gapeq}
\zgap = \pm h \log ( 1 \mp e^{-\zgap} ) 
\eeq
where $h$ is the dimensionless parameter:
\beq
\label{h}
h = \int \dkline  ~ G(0 , \kvec ) 
\eeq
Henceforth, by gap equation we will mean the algebraic equation
(\ref{gapeq}) rather than the integral gap equation (\ref{pseudo}). 
The interactions are encoded in the parameter $h$ which we will
refer to as {\it interaction parameters}.

\begin{figure}[htb] 
\begin{center}
\hspace{-15mm}
\psfrag{A}{$\vep/T$}
\psfrag{b}{$k$}
\psfrag{c}{$\sim k/T$}
\psfrag{d}{$g$}
\psfrag{e}{$\sim T$}
\includegraphics[width=12cm]{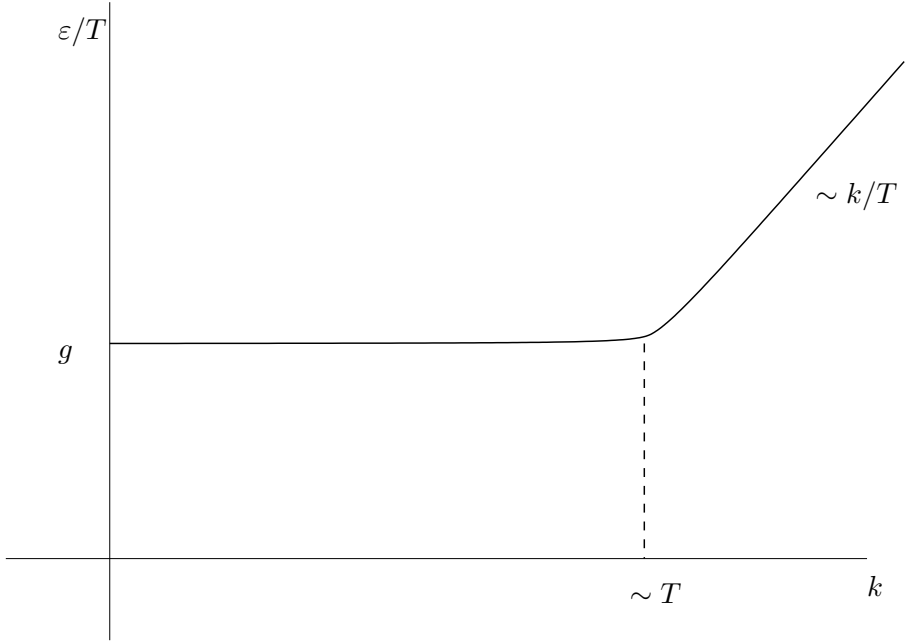} 
\end{center}
\vspace{-2mm}
\caption{The thermal gap.} 
\vspace{-2mm}
\label{Figure1} 
\end{figure} 

The quantity $\cfp$ should only depend on the gap parameter $\zgap$. 
It can be computed to a very good approximation,  if not exactly,
as follows.   In eq. (\ref{4.2}) we wish to trade the integral
over $\kvec$ for an integral over $\vep$.   Differentiating 
eq. (\ref{4.1}) with respect to $k$ 
one finds
\beq
\label{4.3}
k = k\d_k \vep ~ \mp ~  \inv{\beta} k \d_k G*L 
\eeq
where we have defined $L= \log (1\mp e^{-\beta\vep})$.  Inserting
this into eq. (\ref{4.2}): 
\beq
\label{4.4}
\cfp = \mp \frac{\beta^d}{a_d} \int 
\dkline \( k\d_k \vep \mp \inv{\beta} k\d_k G*L \) L(\kvec) 
\eeq
Using the property eq. (\ref{4.5})
and  a few integrations by parts 
one obtains:
\beq
\label{4.6}
\cfp \propto \int_\Delta^\infty
 d\vep ~ k^{d-1} \( \pm L(\vep ) + \frac{\beta \vep}{
e^{\beta \vep} \mp 1 } \)
\eeq
When $\vep > \Delta$, $k\approx \vep$, so we approximate 
$k$ by $\vep$ in the above equation. 
The final result is:
\beq
\label{cuv}
\cfp (\zgap) =
\inv{\Gamma(D-1) \zeta (D) }~ \CL_{D} (\zgap) 
\eeq
where
\beq
\label{Ld}
\CL_{d+1} (\zgap) \equiv \inv{(d+1)} \int_\zgap^\infty dx 
\( 
\mp x^{d-1} \log (1\mp e^{-x} ) + \frac{x^d}{e^x \mp 1 } \)
\eeq
(We have used
 $\int d\kvec = \Omega_d  
\int k^{d-1} dk $, $\Omega_d =(2\pi^{d/2}/\Gamma(d/2))$, and the duplication
formula for the $\Gamma$-function.)
\def\Li{{\rm Li}}

The above integral can be expressed in terms of the 
the polylogarithm, or Jonqui\`ere,  function:
\beq
\label{Jon}
\Li_n (z) = \sum_{k=1}^\infty \frac{z^k}{k^n}
\eeq  
For $D=2,3,4$ one finds:
\barray
\label{Ls}
\CL_2 (\zgap) &=& \pm  \frac{\zgap}{2} \Li_1 (\pm e^{-\zgap} )  \pm \Li_2 (\pm  e^{-\zgap})
\\
\nonumber
\CL_3 (\zgap)  &=& \pm  \frac{\zgap^2}{3} \Li_1 (\pm e^{-\zgap}) 
\pm \zgap \Li_2 (\pm e^{-\zgap}) \pm \Li_3 (\pm e^{-\zgap}) 
\\
\nonumber
\CL_4 (\zgap)  &=& \pm  \frac{\zgap^3}{4} \Li_1 (\pm e^{-\zgap}) 
\pm \zgap^2 \Li_2(\pm e^{-\zgap}) \pm 2\zgap \Li_3(\pm e^{-\zgap}) 
\pm 2 \Li_4 (\pm e^{-\zgap}) 
\earray
Note that $\Li_1 (z) = - \log (1-z)$.

\def\Lr{{\rm Lr}}
\def\cD{c_D}
Since the argument of the 
polylogarithms is always $\pm e^{-g}$, it is convenient
to define the gap variable:
\beq
\label{zgap}
z \equiv e^{-g}
\eeq
To simplify further the subsequent notation, introduce the functions:
\beq
\label{Lrogers}
\Lr_n (z) = \Li_n (z) + \sum_{r=1}^{n-2} \frac{(-)^r}{r!}
 \log^r |z| \Li_{n-r} (z) 
+ (-)^n \frac{(n-1)}{n!} \log^{n-1} |z| \log (1-z) 
\eeq
Comparing with eq. (\ref{Ls}), one finds 
\beq
\label{rat1}
\CL_{d+1} (g) = \pm  \Gamma (d) \, \Lr_{d+1} (\pm e^{-g}) ,  
\eeq
One then has the simple formula for real $z$: 
\beq
\label{cLr}
\cfp  (z) = \inv{\zeta (d+1)}  ~ s   \Lr_{d+1} (s z ) \equiv \cD (z,s)
\eeq
where as before, $s=+/-$ corresponds to bosons/fermions.

For many particles the  gap equation becomes
\beq
\label{many4} 
g_a = \sum_b h_{ab} ~ s_b \log ( 1- s_b e^{-g_b} ) 
\eeq
where 
\beq
\label{many5}
h_{ab} = \int \dkline  ~ G_{ab} (0 , \kvec )
\eeq
In terms of the $z_a$, 
\beq
\label{gapz}
z_a = \prod_b   \( 1- s_b z_b \)^{- h_{ab} s_b }
\eeq
The quantity 
$\cfp$ then becomes a sum:
\beq
\label{many6}
\cfp = \sum_a ~ \cD (z_a , s_a ) 
\eeq

\def\cl{{\rm cl}}

Let us summarize the results of this section.  
The field theory has been deconstructed into the interaction parameters
 $h_{ab}$, which determine the thermal gap parameters $z_a$
through the gap equation.   
The fixed point value of the central charge  $\cfp$ is then
a function of the $z$'s.

 Since $h_{ab}$ determine
the fixed point properties, they  must be an RG invariant functions
of the couplings.   For the  2D  integrable theories, 
it is well known how the $h$'s 
arise from coupling constants in the exact S-matrix. (See section XB.) 
Since we have relaxed the integrability,  let us illustrate how
to compute $h$ to lowest order in the $\phi^4$ theory in 2D, 
which is not integrable. 
The kernel $\Gkern$ was
determined to lowest order in section VI.  The interaction
parameter $h$ to lowest order is then:
\beq
\label{inth1}
h = \frac{\lambdadot}{8m} \int \frac{d\kvec}{2\pi} \inv{\omk^2} 
\eeq
  The above integral is convergent, and using
$\lambdadot = 2 \lambda + ...$, to lowest order one finds
\beq
h = \frac{\lambda}{8 m^2} 
\eeq
Since $\lambda$ has dimension $2$, the ratio $\lambda/m^2$ is
a dimensionless coupling constant, and thus so is $h$.
The $\phi^4$ model may be considered as the weak coupling limit
($b \to 0$) limit of the sinh-Gordon model defined in eq. (\ref{sh.1})
where $h\propto b^2$.   However the integrable sinh-Gordon model
actually has $h=1$ indepependent of $b$ 
(see below) which shows the importance of the 
higher order corrections.      

For higher dimensions it is not clear how the structure of the
kernel $\Gkern$ can lead to finite RG invariant $h$'s,  
and this will remain an important unanswered question in this paper.

\section{Classical statistical mechanics: Ising-like models}

In this section we consider the application of the above formalism
to classical statistical mechanics in $D$ spacial dimensions.  
Since the physical context here is very different, the subscript
${~}_{\rm cl}$ will be used to denote quantities in classical
statistical mechanics to avoid the likely confusion.  The meaning
of $\cfp$ in this context is the following.  Consider a statistical
mechanical system on a $D$ dimensional lattice where all dimensions
of the lattice are infinite except for one of length $L$, i.e. the 
volume of the lattice is $V L$ where $V$ is a $D-1$-dimensional volume
and $V\to \infty$.  At the critical classical temperature $T_c$, 
the partition function should only depend on $L$.  If periodic
boundary conditions are imposed in the finite size direction, 
then $L$ is the same as a finite inverse temperature $\beta$. Thus
the results of section III imply:
\beq
\label{stat}
\lim_{V\to \infty} \inv{V} \log Z_{\rm cl} = \cfp \,  a_{D-1}\,  L^{1-D} 
\eeq

Consider an Ising-like  model on a $D$ dimensional square lattice. 
At each site $i$  there is a spin degree of freedom $\sigma_i = \pm 1$. 
The partition 
function and classical energy $E$ are
\beq
\label{cl1}
Z_\cl = \sum_{\rm config.} e^{-\beta_\cl E } ,~~~~~~
E \equiv J \sum_{<i,j>} \sigma_i \sigma_j 
\eeq
 where $<i,j>$ denotes nearest neighbors, and $\beta_\cl = 1/T_\cl$. 
For $D=2$ the usual Ising model is equivalent to a relativistic
quantum field theory of a free Majorana fermion in euclidean 
$2D$.  Though this is well-known, it is very
non-trivial and relies on mappings to the XY spin chain 
and a Jordan-Wigner transformation. 

No higher dimensional 
version of this is known. We can  explore the possibility 
of a higher dimensional version using the gap equation.
In a disordered phase,  since the spins are disordered, 
one expects that $< E > = 0$.  From the first law of thermodynamics,
one then obtains that the free energy is proportional
to the entropy: $F_\cl = - S_\cl /\beta_\cl$. Now suppose that
at the critical point $\log Z_\cl$ is equivalent to a field theory
functional integral at an ultra-violet fixed point. Let us
clarify that  the temperature
of the field theory $T$ and the classical temperature $T_\cl$ have
nothing to do with each other.   
As is well-known in the 2D case,  the classical temperature appears
as a coupling constant in the zero temperature field theory, the mass of
the fermion being proportional to $T_\cl - T_c$ where $T_c$ is the
critical temperature.  Using the results of section VII, 
one finds in the ultra-violet limit of the field theory for the 
entropy:
\beq
\label{cl2}
S  \approx     \pm V \int \frac{d\kvec}{(2\pi)^d} ~ \log ( 1\mp e^{-g} )
\eeq
where $g$ is the thermal gap.  Since $V/(2\pi)^d$ is approximately
a level density, then 
$
S \approx \pm \sum  \log ( 1\mp e^{-g} )
$.
From this equation, one sees that $g = \log 2$ in the bosonic case
 is
special since then the entropy resembles $S = \sum  \log 2$,
which  suggests a model with 2 states per site, such as an
Ising model.

Though the above discussion is very rough, 
$g=\log 2$ indeed leads to the same  central charge 
$\cfp = 1/2$ as for the  2D Ising model (\ref{cuv}).   For $3D$,    
the same formulas again give a rational result, $\cfp = 7/8$.  
This rationality is quite remarkable given the complexity of the
various functions involved. 
The existence of rational theories will be explored in  the next section.
 In $4D$, rationality ceases and one finds 
$\cfp = \Lr_4 (1/2)/\zeta (4) = 0.97805427...$.  
Clearly more work is needed in the 3D case,  for example
one could  check whether for an infinite slab of thickness $L$,  
$\log Z_{\rm cl} /V$  scales  as  $7 \zeta (3)/16\pi L^2$.

\def\Lr{{\rm Lr}}

\section{Rational theories}

In 2D, theories with a rational $\cfp$ are of particular importance.
For instance unitary theories with $0<c<1$ have been classified
and are the so-called minimal models with rational  
$c=1-6/(n+2)(n+3)$, $n=1,2,...$\cite{BPZ,FMS}. 
Also, all known integrable scattering theories lead to 
thermodynamic Bethe ansatz   
equations with a rational $\cfp$ in the ultra-violet limit. 

 The function $\Lr_2 (z)$ is the Roger's 
dilogarithm.  Remarkably,  the higher order functions $\Lr_n$
have also appeared in the mathematical literature\cite{Lewin}.
They were not originally defined in terms of the integral 
of section VII.  
Rather, these functions arose as the unique functions that remove 
logarithmic terms in certain functional equations satisfied by
$\Li_n (z)$. 
We can now formulate the following problem.  A rational theory
in $D$ dimensions 
requires the following identity:
\beq
\label{SPL}
\sum_{a=1}^N  s_a \, \Lr_{D} (s_a z_a ) =  \CR ~ \zeta (D)
\eeq
where $s_a = \pm 1$ corresponds to the statistics and 
$z_a$ are numbers that characterize the thermal gaps $g_a$, 
$z_a = e^{-g_a }$. 
If $\CR$ is rational, then 
$\cfp = \CR$ is rational. 
We will refer to a relation of the form eq. (\ref{SPL}) as
a Rational Statistical Polylogarithmic Evaluation (RSPE).
Every RSPE may be viewed as a non-trivial result in  number
theory.

Of course, a  relation of the form eq. (\ref{SPL}) does not guarantee 
the existence of a  field theory with this $\cfp$
in the UV.  However one can try to go as far as possible
in reconstructing some basic data of a possible theory.
 Given the
$z_a$ from an RSPE,  one can 
 use the gap equation to determine the interaction parameters $h_{ab}$. 
 There 
then  remains the difficult problem of finding 
a consistent  field theory leading to kernels 
$\Gkern_{ab}$ that yield the  $h_{ab}$.  
In 2D, if an
integrable theory exists with $\cfp$, the theory can often
be entirely reconstructed, i.e all the scattering amplitudes can
be deduced,  due to the tight constraints of integrability.   
The  encouraging fact is that, as we show below, 
we have been able to reconstruct the  field theories for all known one-particle (N=1) 
RSPE's  in 2D.                 
We have also been able to reconstruct a 2-particle theory
in $2D$.

  The
classification of fixed points of field theories is then
closely related to the classification of RSPE's.  
An RSPE is a subclass of the more general polylogarithmic ladders
 constructed  in \cite{Lewin}.   The latter do not have
the restriction that the statistical parameters $s_a$ enter the
relation as in (\ref{SPL}),  though some polylogarithmic ladders
 can be brought
into the form of RSPE's using certain functional transformations
(see below).  The gap equation (\ref{gapz}) is essentially
what is called the cyclotomic equation in the construction
of polylogarithmic ladders.   
The forerunners of dilogarithmic ladders (d=1) were first found
by Euler and Landen in the 1700's\cite{Euler,Landen}. 
New dilogarithmic ladders
were not found until the 1930's by Coxeter and Watson.  In the
1980's it was realized by Lewin that there is a vast extension 
of these previous few results.   Despite this progress,  
polylogarithmic ladders are still considered somewhat mysterious
and there are no classification theorems.  Some of them have only
been verified to high precision numerically.  Ladders have even
been discovered to order $D=17$\cite{Order 17}.   In the sequel
we will present a number of RSPE's that were deduced from
results in \cite{Lewin,Lewin2,Abou}, although they do not all appear
there in the form they are given here;  the diligent reader
can easily verify them numerically (as we did).

Let us first describe the only results known that are valid
for all dimension $D$.  Polylogarithmic ladders are constructed using
functional relations satisfied by $\Lr_n (z)$, without doing
any explicit integrals. 
For any $n$, $\Lr_n (1) = \zeta (n)$.  This corresponds to 
a free boson in any dimension with $\cfp = 1$.   The functions
$\Lr_n (z)$ satisfy a number of functional equations that are valid
for any $n$.   
The duplication relation reads:
\beq
\label{ladder1}
\Lr_n (z) + \Lr_n (-z) = 2^{1-n} \Lr_n (z^2 ) 
\eeq
From this it follows that $- \Lr_{d+1} (-1) = (1-2^{-d}) \zeta (d+1)$.  
This corresponds to a free fermion in any dimension with 
\beq
\label{cfer}
c_{d+1} (1, -1)  = 1- \inv{2^d} , ~~~~~~~~~({\rm free ~ fermion})
\eeq

For any $n$ one also has the inversion relation:
\beq
\label{inv}
\Lr_n (-z) + (-)^n \Lr_n (-1/z ) = C_n 
\eeq
where $C_n $ is a constant.  For $n$ odd, $C_n = 0$,
whereas for $n$ even 
$C_n = -2(1-2^{1-n}) \zeta (n)$.
Using $\Lr_n (0) = 0$ in the inversion relation, one finds:
\beq
\label{hard}
c_{d+1} (\infty , -1 )  
 = 2 \( 1- \inv{2^d} \) 
~~~~~~( d ~ {\rm odd}) 
\eeq
The above relation is fermionic, and since $z=\infty$,  
it is an interacting theory with gap $g=-\infty$ and
interaction parameter $h=1$.  
As we explain in the next subsection,  in $2D$ this theory
is realized in e.g. the sinh-Gordon model with $\cfp = 1$. 
In higher dimensions  with $d$ odd, $\cfp = 2(1-2^{-d})$ which is
twice that of a free fermion and  always more than 
a boson.  We will refer to this case as the  extreme-fermion
since it possesses the highest possible $\cfp$ for a fermionic
theory.

Notice that the above results were obtained without doing any
integrals explicitly, but rather just by using functional relations
satisfied by the $\Lr_n (z)$.  More complex polylogarithmic
ladders are constructed in the same spirit.  
Since the more recently discovered ladders are for $N>1$, 
a classification of 1-particle theories is essentially  within reach. 
In the next section we carry out this classification.

\section{Rational theories in  $2D$}

\subsection{One particle RSPE's}

Let us illustrate how ladders are constructed in the simple case of
$N=1$ (1-particle) and $D=2$.   The construction relies on additional
 functional relations which are valid at  second order:
\barray
\nonumber
&\Lr_2 (z)& + \Lr_2 (1-z) = \zeta (2) 
\\
\label{lad2}
&\Lr_2 (z)& + \Lr_2 ( -z/(1-z) ) = 0
\earray
Using the first relation above, one sees that
$\Lr_2 (1/2) = \zeta (2)/2 $.   This, and the free fermion and
free boson cases,  were the relations
known to Euler.   Landen found more evaluations as follows.
If $r$ is a root to the polynomial equation
$z^2 + z  = 1$,  then the above functional equations
become linear equations for $\Lr_2 (z)$ with argument
$z=r, -r, -1/r$ and $r^2$. 
Let us take $r=(\sqrt{5} -1 )/2$, which is the inverse of the
golden ratio.  This is the  complete list of known RSPE's for $D=2$
and $N=1$, and has been known for over 220 years:
\barray
\nonumber
\Lr_2 (1) &=& - \Lr_2 (-\infty ) = \zeta (2) 
\\
\label{euler}
\Lr_2 (1/2 ) &=& - \Lr_2 (-1) = \inv{2} \zeta (2) 
\\
\nonumber
\Lr_2 (r^2 ) &=& - \Lr_2 (-r) = \frac{2}{5} \zeta (2) 
\\
\nonumber
\Lr_2 (r ) &=& - \Lr_2 (-1/r) = \frac{3}{5} \zeta (2)
\earray
Though it is generally believed that the above list is
exhaustive,  
we could not find  a proof in the literature.    
We suggest that a proof might be concocted based on 
proving that the known integrable scattering theories in 2D
with one particle are exhaustive and are in correspondence
with the above list.  

Assuming that the relations eq. (\ref{euler}) are indeed exhaustive, 
this leads to the classification 
of the  possible theories shown in
Table 1.  In the table are listed the thermal gaps $g$ and
the interaction parameter $h$ which can be inferred from
the gap equation (\ref{gapeq}).  The last 3 columns will be
explained below.

\bigskip

\begin{center}
\begin{tabular}{|c|c|c|c|c|c|c|}
\hline
statistics & $\cfp$ & $g$ & $h$ & $S(0)$ &$s_f$  & field theory
 \\
\hline 
fermion & 1 & $-\infty$ & 1 & $-1$ & 1& sinh-Gordon   \\
boson & 1 & 0  & 0 & 1 & 1&  free boson   \\
fermion & 1/2 & 0 & 0 & 1 & $-1$ & free fermion   \\
boson & 1/2 & $\log 2$& $-1$ & 1 & 1 &  \\
fermion & 2/5 & $-\log r$  & $-1$ & $-1$ & 1 & Lee-Yang   \\
boson & 2/5  &  $-\log r^2$ & $-2$ & 1 & 1  &     \\
fermion & 3/5  & $-\log 1/r$  & $1/2$  & 1 & $-1$  &  $\CM_{3,5}$ \\
boson & 3/5  & $-\log r$& $-1/2$  & 1 & 1 &   \\
\hline
\end{tabular}
\end{center}
\bigskip

\subsection{Integrable reconstruction in 2D}

Let us specialize the results of section V  to 
integrable  relativistic theories in $2D$\cite{ZamoTBA}.
  Let $A_a (\theta )$,
$a=1,..,N$ denote particle creation operators where $\theta$ is
the rapidity parameterizing the one-particle relativistic dispersion relation:
\beq
\label{1d1}
E = m_a \cosh \theta , ~~~~~~\kvec = m_a \sinh \theta
\eeq
In this rapidity space the formulas  for
the partition function and
integral gap equation take the  form:
\beq
\label{1d3}
\log Z = -V \sum_a  \int_{-\infty}^\infty \frac{d\theta}{2\pi} 
 m_a \cosh\theta s_a \log (1-s_a e^{-\beta \vep_a (\theta )} )
\eeq
\beq
\label{1d4}
\vep_a (\theta ) = (m_a \cosh \theta - \mu_a )  + \inv{\beta} \sum_b  
\int_{-\infty}^\infty \frac{d\theta'}{2\pi}
~ G_{ab} (\theta - \theta' ) s_b \log (1-s_b e^{-\beta \vep_b (\theta' )} )
\eeq

We will consider only theories where the scattering is diagonal. 
The factorizable S-matrix $S_{ab} (\theta ) $ then  enters the exchange
relation as follows:   
\beq
\label{1d2}
A_a (\theta ) A_b (\theta') = S_{ab} (\theta - \theta' ) 
A_b (\theta' ) A_a (\theta)
\eeq

In the case of $2D$ an alternative derivation of the kernel
is possible\cite{Yang,ZamoTBA}.
Consider the quantum field theory on a space of 
1 dimensional length $L (=V)$.  Requiring that the multi-particle
wave-function is periodic leads to a quantization condition 
on the momenta $\kvec$:
\beq
\label{1d6}
e^{iL \kvec_i } ~ \prod_{j\neq i} S_{ij} (\theta_i - \theta_j ) = 1
\eeq
Taking the logarithm and differentiating with respect to the rapidity
leads to a relation between $\rho_l $ and $\rho$ as in eq. (\ref{3.2})
with the kernel:
\beq
\label{1d5}
G_{ab} (\theta ) = -i \dd{~}\theta  \log S_{ab}
\eeq

This derivation has its origins in the thermodynamic Bethe ansatz (TBA).  
In many cases one can construct explicitly the Bethe ansatz
wave-functions and derive equations such as eq. (\ref{1d6}).    
For this reason we will refer to the above 2D equations as 
TBA equations.   

On the other hand, using the arguments of the last section, 
the  kernel $\Gkern$ can be determined from the form factors. 
It was shown in \cite{LecMuss} that if one defines the connected
form factor in the proper way, then the resulting kernel is
the same as in eq. (\ref{1d5}).

We now attempt to 
identify the entries of the above table with specific integrable 
quantum field theories.  All of the fermionic theories are well-known
and their conventional names are in the last column.    However each 
fermionic
theory has a bosonic counterpart, which can be traced to the
second functional relation eq. (\ref{lad2}).   As explained in
\cite{ZamoTBA}, the statistics of the lagrangian fields does
not necessarily correspond to the statistics in the partition
function.  First of all, there is no spin-statistics theorem
in 2D.  The S-matrix satisfies the unitarity condition
$S(\theta ) S(-\theta ) = 1$.  At $\theta = 0$, 
$S(0)^2 = 1$ has two possibilities: $S(0) = \pm 1$.   
The quantization condition eq. (\ref{1d6}) imposes its own
restrictions on the wave-functions.  If $S(0) = -1$, two
bosons are not allowed to have the same rapidity, and one
should impose fermionic exclusion statistics.  If $S(0) = 1$,
the situation is reversed.  We can summarize this as follows.
Let $s_f = \pm 1$ denote the statistics of the fields, where
$s_f = 1$ is a boson.  Let $s$ denote the statistics of the partition
function, as in eq. (\ref{1d3}).  Then $s = s_f S(0)$.  
Henceforth, in the 2D case, boson or fermion statistics will
refer to the partition function statistics $s$.  

The reason that half the theories in Table 1 have not been identified  
is
that nobody has ever made sense of an interacting {\it bosonic} TBA.
It has even been suggested that interactions always turn bosons
into fermions in 2D\cite{Klassen}.    
This issue was addressed in \cite{Mussbos}, where a hypothetical
model with S-matrix given by the well-known sinh-Gordon one, 
but with the wrong statistics,  was investigated. 
 It was found that
$\cfp$ was ill-defined in the UV, which shows one cannot simply
change the statistics of known theories at will.   On the other hand, 
as Table 1 shows,  it is possible to have well-defined $\cfp$
for a bosonic TBA.  

For the remainer  of this section we describe how to reconstruct S-matrices
from the data  in Table 1.   
We first review a few basic ingredients of diagonal
factorizable S-matrices (see for instance \cite{ZZ,Klassen,FateevZ}).
   Unitarity and crossing symmetry
require
\beq
\label{cross}
S_{ab} (\theta ) S_{ba} (-\theta ) = 1, ~~~~~~~~~
S_{a\bar{b}} (\theta ) = S_{ab} (i \pi - \theta )
\eeq
where 
$\bar{a}$ is the charge conjugate to $a$.  S-matrices satisfying
the above equations are products of the fundamental building blocks:
\beq
\label{fss}
f_\alpha (\theta ) = \frac{ \sinh \inv{2}(\theta + i \pi \alpha) }    
{\sinh\inv{2}(\theta - i \pi \alpha)}
\eeq
If a particle is its own anti-particle, then crossing symmetry
implies it is a product of the functions
\beq
\label{Fss}
F_{\alpha} (\theta ) = f_\alpha (\theta ) f_\alpha (i\pi - \theta )
= \frac{ \tanh \inv{2}(\theta + i \pi \alpha) }    
{\tanh\inv{2}(\theta - i \pi \alpha)}
\eeq

The poles in the S-matrix are also severely constrained.  If a simple  pole
exists in $S_{ab}$ at $\theta = i u$,  on the physical strip
$0<iu<i\pi$, and with
positive imaginary residue,  then this
corresponds to a stable bound state,  with 
mass
\beq
\label{boundstate}
m^2 = m_a^2 + m_b^2 + 2 m_a m_b \cos u 
\eeq
If this is a new particle, one must proceed to close the bootstrap.
A negative residue pole is either the same particle in the crossed
channel, or a sign of non-unitarity. 
We will need that $f_\alpha$ has a simple pole at $\theta = i\pi \alpha$
with residue $2i\sin \pi\alpha$.  From this it follows
that $F_\alpha$ has a simple pole at $i\pi\alpha$ with residue
$2i\tan \pi\alpha$ and a simple at $i(1-\alpha)\pi$ with 
residue $-2i\tan \pi\alpha$.

The $h$ values that appear in the gap equation 
for the 1d integrable theories are:
\beq
\label{hab}
h_{ab} = \int_{-\infty}^\infty \frac{d\theta}{2\pi i} ~ 
\dd{ \log S_{ab} (\theta )}\theta 
\eeq
Using this one easily can compute that for each factor of
$f_\alpha$,  $h= \alpha - {\rm sign}(\alpha)$.  For
each $F_\alpha$, $h=-{\rm sign}(\alpha)$.  

All the theories with $h\neq 0$ are interacting.  Let us first
go over the identification of the interacting fermionic theories, all of
which are known.  

\bigskip

\noindent {\bf Fermionic theory with $\cfp = 1$.}  ~~~
Here since $h=1$ the S-matrix must be $F_{-\alpha}$ with
$\alpha$ positive.  For $\alpha$ positive, the pole on the physical
strip has negative residue and thus does not correspond to a new particle,
so the bootstrap is already closed.  This is a one-parameter family
of theories corresponding to the sinh-Gordon model, with lagrangian:
 \beq
\label{sh.1}
S_{\rm shG} =  \int d^2 x  ~\(  \inv{2} (\d \phi)^2 + 
\Lambda \cosh ( b \sqrt{4\pi} \phi )  \) 
\eeq
The S-matrix is known to be $F_{-\alpha}$ where 
$\alpha = b^2 /(2+b^2)$.  $\phi$ is a bosonic field, so $s_f = 1$,
but since $S(0) = -1$, the partition function is fermionic.

\bigskip
\noindent {\bf Fermionic theory with $\cfp = 2/5$.}  ~~~
Since here $h=-1$, the S-matrix must be $F_\alpha$ with
$\alpha$ positive.   However for arbitrary $\alpha >0$, 
the physical pole will correspond to a new particle
and this is no longer a one-particle theory.  The only solution
is to take $\alpha = 2/3$.  Then, since $\cos 2\pi/3 = -1/2$,
the  pole corresponds to a particle of the same
mass as the original particle and can thus be identified with it.
Thus the bootstrap closes with one particle.   This model is
known as the Lee-Yang  model\cite{YangLee}. 
Generally, $\cfp = c - 24\Delta_{\rm min}$
where $c$ is the Virasoro central charge and 
$\Delta_{\rm min}$ is the lowest conformal dimension of the fields.
Here, $c= -22/5$ and $\Delta_{\rm min} = -1/5$,
giving $\cfp = 2/5$. The residue of the pole is negative,
which is an indication of the non-unitarity of the theory.

\bigskip
\noindent {\bf Fermionic theory with $\cfp = 3/5$.}  ~~~
Here since $h=1/2$ the S-matrix must be $f_{-1/2} 
= -i \tanh \inv{2} (\theta - i\pi/2 )$.  Though $f_\alpha$
for general alpha cannot correspond to a 1-particle theory
since it is not crossing symmetric, $\alpha =- 1/2$ is 
special since it is at least anti-crossing symmetric:
$f_{-1/2} (i\pi - \theta ) = - f_{-1/2} (\theta )$. 
This model is known\cite{M35}  to be the $\Phi_{13}$ perturbation of
the $\CM_{3,5}$ minimal model, with $c=-3/5$ and 
$\Delta_{\rm min} = -1/20$, giving $\cfp = 3/5$.

\bigskip

We now turn to the bosonic theories. Note that they all occur
at the same central charge as the fermionic ones.  This does not
imply they are the same theory, but only that they are the
same at the UV fixed point.  A theory is defined by its fixed
point in the UV and by the operator which perturbs it away
from the fixed point\cite{ZamoE8}. For the last 4 entries of
Table 1, we only indicate the UV fixed point theory.

\bigskip
\noindent {\bf Bosonic  theory with $\cfp = 1/2$.}  ~~~
This is the most interesting case since the UV central charge is
that of the Ising model, of which a great deal is already known. 
Let us attempt to reconstruct a consistent  S-matrix for this theory deferring
its interpretation. 
   Since $h=-1$, the 
S-matrix must be $F_\alpha$ with $\alpha$ positive. For
the same reasons given above for Lee-Yang, $\alpha$ must
be $2/3$ otherwise the bootstrap doesn't close.
We propose the following S-matrix: 
\beq
\label{isingS}
S (\theta ) = p  F_{2/3} (\theta) ,  ~~~~~p=\pm 1 
\eeq

We now carefully investigate the consistency of this
theory.  For $p=-1$,
the residue of the pole at $\theta = 2\pi i/3$ is positive,
which is consistent with the theory being unitary. 
However the  S-matrix now satisfies the bootstrap equation  with
an extra minus sign:
\beq
\label{bootising}
S(\theta ) = p S(\theta + i \pi/3 ) S(\theta -i\pi/3 ) 
\eeq
This minus sign can  be removed by $S \to -S$,
however this spoils the unitarity of the theory. 
For these reasons, the above S-matrix is not entirely
consistent and its  physical interpretation 
 will remain an open question.  The same is true for the next two
bosonic theories.

\noindent {\bf Bosonic  theory with $\cfp = 2/5$.}  ~~~
Here $h= -2$ which implies the S-matrix must be
$F_\alpha F_{\alpha'}$. As in previous cases, the bootstrap
will not close unless $\alpha = \alpha' = 2/3$:
\beq
\label{anS}
S (\theta ) = F_{2/3} (\theta )^2
\eeq
  The resulting
S-matrix has only double poles for the physical particle.  Since
$S(0) = 1$, this model has a single bosonic field.  Further analysis
is needed to determine whether it is equivalent to the Lee-Yang model,
or  a different off-critical perturbation.  

\bigskip
\noindent {\bf Bosonic  theory with $\cfp = 3/5$.}  ~~~
Since $h=-1/2$, the S-matrix must be:
\beq
\label{addit}
S(\theta ) = - f_{1/2} (\theta )
\eeq
It has a pole at $i\pi / 2$, with negative residue,
indicating non-unitarity. Here, $s_f = 1$.    
As for the fermionic version, 
this is an integrable perturbation of the $\CM_{3,5}$ 
non-unitary minimal model.  Since the S-matrices
are very similar in this case for the bosonic and fermionic
theories ($f_{1/2} (\theta ) = f_{-1/2} (-\theta )$),
the two models are probably equivalent;  again more analysis 
is needed.

\subsection{A supersymmetric 2 particle theory in  $2D$}

As was stated before, all of the newly discovered ladders
are multi-particle.  Not all of them can be brought into the
RSPE form.  It is beyond our scope to do a comprehensive study
of the vast number of results in \cite{Lewin}, so we concentrate
on a simple example that doesn't appear to correspond to anything 
already known.  Using the duplication and inversion relations 
on one of the results in \cite{Lewin}, one can show 
\beq
\label{susy}
c_2 ( y , 1) + c_2 (y, -1 ) = 
  \frac{3}{4} 
, ~~~~~~y= \sqrt{2} -1 
\eeq
This is an interesting relation since, unlike the known TBA's
that are purely fermionic,  this is a mixed theory, with one
boson and one fermion, suggesting a supersymmetric theory.  

We can reconstruct the theory just from the data in the
equation (\ref{susy}).  It's a two particle theory with
$z_1 = z_2 = \sqrt{2} -1$, but with $s_1 = 1 , s_2 = -1$.  
From the gap equation we determine the $h$'s to be:
$h_{11}= h_{22}= h_{12} = h_{21} = -1$.  The choice of
S-matrix:
\beq
\label{Ssusy}
S_{ab} (\theta ) = - F_{2/3} (\theta ) , ~~~~~~~~
\forall a,b = 1,2 
\eeq
yields these $h$ values, and the bootstrap is closed for
previously described reasons.  The S-matrix shows no 
signs of non-unitarity.   However the  UV central charge is 
$c=3/4$ which is not in the minimal unitary series. 
Our  interpretation is that the physically sensible theory is
two copies of the above.  This has $c=3/2$ with 4 particles
and the supersymmetry is preserved.  It is known that $c=3/2$
is the highest central charge of minimal unitary super-conformal 
series\cite{susymin},
supporting this idea.

\section{RSPE's  in  $3D$ and $4D$. }

\subsection{One particle theories in $3D$}

As explained in the last section,  the second identity in
eq. (\ref{lad2}) implies that in 2D  a bosonic RSPE can be 
rewritten as a fermionic one with different gap parameter $z$. 
Because of bosonization,  this is perhaps not surprising in 2D. 
In higher dimensions there is no analog of this identity 
so it is not possible in general to give both a bosonic
and fermionic interpretation to a given RSPE. 

The only known 1-particle RSPE's  (due to Landen\cite{Landen})
beyond the free boson and  free fermion in $3D$  
have again been known for over 220 years and are the following:
\beq
\label{2d1}
c_3  \( \inv{2} , 1 \)  = \frac{7}{8}  , ~~~~~
c_3 (r^2, 1) =  \frac{4}{5}  
\eeq
where again $r= (\sqrt{5}-1)/2 $.  
This leads to the following table of possible 1-particle rational theories
in $3D$ which is again very likely to be exhaustive, but we cannot 
prove it.

\bigskip

\begin{center}
\begin{tabular}{|c|c|c|c|}
\hline
statistics & $\cfp$ & $g$ & $h$ \\ 
\hline 
fermion & 3/4  & $0$  & $0$  \\ 
boson   & 1    & $0$  & $0$  \\  
boson & 7/8  & $\log 2$  & $-1$  \\ 
boson & 4/5   & $-\log r^2$  & $-2$  \\ 
\hline
\end{tabular}
\end{center}
\bigskip

The interacting bosonic theory with $\cfp = 7/8$ was suggested to be
the 3D Ising model in section VIII.   The $\cfp = 4/5$ theory appears to be 
related to 
the $O(N)$ sigma model, 
which at large $N$ has $\cfp = \frac{4}{5} N$\cite{Sachdev}.
The method for computing $\cfp$ in the latter work is just  the effective
action at large $N$, and is thus different than our methods.

\subsection{Multi-particle RSPE's  in $3D$}

For the remaining RSPE's presented in this paper we only
reconstruct the particle spectrum, gap equation, and 
interaction parameters $h_{ab}$, since as previous stated
we do not have the tools necessary to further reconstruct
the theory.   There are a number of generic features of these
multi-particle RSPE's.  They are generally of mixed statistics,
i.e. involve both bosons and fermions.   One can also check
that in these examples, 
the interactions always reduce $\cfp$ from it's free field value,
i.e. $\cfp <  N_b + (1-1/2^d) N_f $, where $N_{b,f}$ are the numbers
of bosons and fermions. Below, we present this feature with the
parameter $r = \frac{\cfp}{c_{\rm free}}$.  
 However this is not generally the case,
as interactions in fermionic theories can increase  $\cfp$, 
the extreme-fermion being the limiting case. (See eq. (\ref{ranges}).) 
 RSPE's appear to be more and more scarce as one moves up in dimension.
This is 
not surprising since conformal field theories are also expected
to be more scarce since there is no known underlying Virasoro algebra
with a rich representation theory.

Let us reconstruct the following $\cfp = \frac{13}{6}$ RSPE:
\beq
\label{rspe3d1}
2 c_3 (1/3 , 1) + c_3 (1/3, -1) = \frac{13}{6} 
\eeq
It contains 2 bosonic particles with $z_1 = 1/3$ and
one fermionic particle with $z_2 = 1/3$.  Here $r=26/33$. 
  The gap equations
then read
\beq
\label{rspe3.2}
z_1 = (1-z_1)^{-2h_{11}} (1+z_1)^{h_{12}} 
,
~~~~
z_2 = (1+z_2)^{h_{22}} (1-z_1)^{-2h_{21}}
\eeq
The solution is $h_{ab} = -1, ~~\forall \, a,b$.  

\bigskip

Consider next the $\cfp = \frac{11}4$ RSPE with 3 bosons and 2 fermions
with $r=11/18$.  
\beq
\label{rspe3d2}
3 c_3 (x,1) + 2 c_3 (x, -1 ) = \frac{11}{4}  
, ~~~~~x \equiv 3 - 2\sqrt{2}
\eeq
Here the gap equation reads:
\beq
\label{rspe3.4}
z_1 = (1-z_1)^{-3h_{11}} (1+z_2 )^{2h_{12}} , 
~~~~
z_2 = (1+z_2)^{2 h_{22}} (1-z_1)^{-3h_{21}} 
\eeq
The novel feature here is that there is a one-parameter family
of solutions to $h$: 
\beq
\label{rspe3.5}
h_{21} = h_{11}, ~~~~h_{12}= h_{22} = 
\frac{ 3 h_{11} \log (1-x) + \log x}{2\log (1+x) } 
\eeq
This suggests some exactly marginal directions since the $h$'s
are required to be RG invariants.

\subsection{Multi-particle RSPE's  in $4D$}

In 4D 
there are no known RSPE's  beyond the free fermion, free boson
and extreme-fermion.  
We present only two examples of multiparticle RSPE's.

Here is an RSPE  with $\cfp = 179/2$ and $r=716/721$: 
\beq
\label{3d1}
42 \, c_4 \( 1/2, 1 \)  +
 c_4 \(  1/8, -1 \)  + 54 \, c_4 (2, -1) = \frac{179}{2} 
\eeq
It consists of 42 bosons and 55 fermions.   
Here  there is a 3 parameter family of interaction parameters  that
give the gap parameters  $z$:
\barray
\nonumber
h_{12} &=& 14 h_{11}, ~~~~h_{21} = \inv{14} (h_{22} -1), ~~~~~
h_{13} = -\frac{14}{27} h_{11} 
\\
\label{4dint}
h_{31} &=& \inv{42} (1-81 h_{33} ) , ~~~~~h_{23} = -\inv{27} h_{22},
~~~~~h_{32} = -27 h_{33}
\earray

\bigskip

We  leave the following  $\cfp = 4393/8$, $r=4393/4513$ 
 example as an exercise: 
\beq
\label{3d2}
264 \, c_4 (\chi , 1) + 15 \, c_4(\chi^2 , -1 )
 + 4\, c_4 (\chi^3 , -1) + 324 \, c_4 (1/\chi , -1) 
= \frac{ 4393}{8} 
\eeq
where $\chi  \equiv 2 - \sqrt{3}$. 

\def\vtheta{\vartheta}

\def\canyon{{\tilde{c}}}

\section{Fractional statistics and imaginary chemical potential}

\subsection{General arguments}

In the formalism we have developed so far, the bosonic
or fermionic statistics $(s=\pm 1)$ has played a crucial role.  
It is well-known that in 2D and 3D, fractional statistics particles,
i.e. anyons, are also possible\cite{anyons}.  In this section 
we argue that this possibility can be incorporated with an 
imaginary chemical potential.  The connection between anyons 
and an imaginary chemical potential has previously been 
suggested in the context of the 3D Gross-Neveu model\cite{Petkou3}.

Let us model our  discussion based on the 3D case, though we will
also present results in 2D.  The standard way to obtain anyons
is to add to the action a topological, or Hopf,  term:
\beq
\label{Anyon1}
S \to S + i \vtheta \, S_{\rm top}
\eeq
The simplest case, which we will refer to as the free anyon,
corresponds to $S$ being free charged bosons or fermions
with $U(1)$ current $J^\mu$ and charges $Q= \pm 1$. 
Here  $S_{\rm top}$ is the Chern-Simons term
where the $U(1)$ gauge field $A_\mu$ is manufactured from the 
$U(1)$ matter current $J^\mu = \epsilon^{\mu\nu\alpha} \d_\nu A_\alpha$.  
The statistical parameter $\vtheta$ is normalized so that
 the spin of the particle is $\vtheta / 2\pi$. 

At finite temperature, space-time has the topology of 
$R^d \otimes S^1$, where $S^1$ is the one dimensional circle.  
Let $\CC$ denote the configuration space of the field theory; 
for example for the $O(3)$ sigma model, $\CC = S^2$.  
At finite temperature the relevant homotopy group is the one
classifying maps from $ R^d \otimes S^1 \to \CC$ which we will
denote as $\pi_T$.  

Let us assume $\pi_T = \Zmath$ and the Hopf term $S_{\rm top}$ is $\pm 1$ for
the particles of $U(1)$ charge $Q= \pm 1$.  This implies that the particles
contribute $\pm i \vtheta$ to $\log Z$.  Comparing with eq. (\ref{3.3}), 
this corresponds to a chemical potential $\mu = \pm i \vtheta \, T$
where $T$ is the temperature.  More generally, let $\mu_a$ denote the
imaginary chemical potential which modifies the statistics of the 
a-th particle with statistical 
parameter $\theta_a$:
\beq
\label{Anyon2}
\mu_a  =  i \vtheta_a  \, T
\eeq
The important feature of this 
{\it statistical chemical potential} is that $ \mu_a/T$ remains
finite at any temperature,  which is a consequence of the fact that
$S_{\rm top}$ is finite.  This leads to a finite contribution to
the thermal gaps $g_a = -i \vtheta_a $.  
In terms
of the gap variables $z_a$ the gap equation reads:
\beq
\label{Anyon3}
z_a = e^{i\vtheta_a } \prod_b (1-s_b z_b )^{-h_{ab} s_b} 
\eeq
where as before $s_a = \pm 1$ corresponds to whether the original
undressed particles were fermions or bosons.

If the particles are free, $h_{ab} = 0$ and $z_a = e^{i\vtheta_a }$, 
i.e. the thermal gaps are purely statistical.  Let us
denote by $\canyon_{D} (\vtheta, s)$  the value of $\cfp$ in D-dimensions  
where $s=1$ ($s=-1$)  corresponds to 
$\vtheta$-statistically dressed charged free bosons (fermions)
with $\vtheta_\pm = \pm \vtheta$.  
 From the results of
section VII, we propose the following formula  for these  ``free
anyons'': 
\beq
\label{Anyon4}
\canyon_D (\vtheta , s ) = \frac{s}{\zeta (D)} 
\( 
 \Lr_D ( s e^{i\vtheta} ) + \Lr_D ( s e^{-i \vtheta } ) 
\) 
\eeq
(For $z$ on the unit circle, $|z|=1$ in the formula (\ref{Lrogers}) 
so that $\Lr (z) = \Li (z)$.)

\subsection{Free anyons in 2D}

A non-trivial check of eq. (\ref{Anyon4}) is available in 2D due
to the  remarkable dilogarithm identities for $z$ on the unit circle: 
\barray
\label{Anyon5}
\canyon_2 (\vtheta, 1 ) =  2 \( 1-  \frac{3 \vtheta}{\pi} 
\( 1-\frac{\vtheta}{2\pi} \) 
\) , ~~~~~ 0 < \vtheta < 2\pi 
\\
\label{Anyon6}
\canyon_2 (\vtheta , -1) = 1- 12 \( \frac{\vtheta}{2\pi} \)^2 
~~~~~ -\pi < \vtheta < \pi
\earray

The arguments leading to eq. (\ref{Anyon4}) assumed the underlying
field theory was a free theory of charged fractional statistics particles.
In 2D the spin $\sigma$ of a field can be defined by monodromy
property:
\beq
\label{spin2d}
\psi ( e^{2\pi i } z , e^{-2\pi i } \zbar  ) = e^{2\pi i \sigma} 
\psi (z, \zbar)
\eeq
where $z,\zbar$ are the euclidean light-cone space-time variables,
$z= x+i y, \zbar = x-i y $. 
The simplest free theories of anyons in 2D  are the so-called $\beta-\gamma$ 
systems\cite{betagamma} which were first studied in connection with
the world-sheet ghosts in string theory.  The theories are defined
by the action 
\beq
\label{Anyon7}
S = \inv{2\pi} \int d^2 x ~ \( \beta \d_\zbar \gamma + \bar{\beta} 
\d_z \bar{\gamma} \)
\eeq 
where the spin of $(\beta, \gamma)$ is $(\sigma, 1-\sigma)$ and the opposite
sign for $(\bar{\beta}, \bar{\gamma})$.   The fields  can be quantized
as bosons or fermions, the latter usually being referred to as
a $b-c$ system.  

Consider first the bosonic $\beta-\gamma$ system.  The central charge
is known to be\cite{betagamma}:
\beq
\label{Anyon8}
c_{\beta-\gamma} = 2 ( 6 \sigma^2 - 6 \sigma + 1)
\eeq
Due to the identity eq. (\ref{Anyon5}), 
this corresponds precisely  to the  statistically dressed boson, 
i.e. $c_{\beta-\gamma} = 
\canyon_2 (\vtheta , 1)$ with $\vtheta /2 \pi = \sigma$.

Consider next the case where $\beta-\gamma$ are quantized as fermions,
usually denoted $b-c$.   
The central charge is known to be:
\beq
\label{Anyon9}
c_{b-c} =   -  2 ( 6 \sigma^2 - 6 \sigma + 1)
\eeq
Since the undressed particles are already fermions, 
by the same reasoning as before, this should correspond to 
a statistically dressed fermion with $\vtheta/ 2\pi = \sigma - 1/2$; 
it is easily confirmed that $\canyon_2 (\vtheta, -1) = c_{b-c}$ using
eq. (\ref{Anyon6}).  

It is somewhat remarkable that we have been able to obtain the
above central charges simply from a statistical thermal gap.
We point out that equality of $c_{\beta-\gamma}, c_{b-c}$ with
$\canyon_2 (\vtheta, s)$ is only valid for the range of $\vtheta$ 
indicated in eqs. (\ref{Anyon5},\ref{Anyon6}) since $\canyon_2 (\vtheta,s)$ 
is a periodic function of $\vtheta$ whereas $c_{\beta-\gamma}, c_{b-c}$
are quadratic.

\subsection{Rational free anyons in 3D}

As the equations (\ref{Anyon5}, \ref{Anyon6}) show, when 
$\vtheta / 2\pi $ is rational then $\canyon_2 (\vtheta, s)$ is 
rational.  In 3D, there is no analog of these equations
and $\canyon_3 (\vtheta, s)$ is generally irrational.  However
for some special values of $\vtheta$, $\canyon_3$ turns out to
be  rational. 

Anyons of spin $1/3$ or $ 1/6$ are  examples of  rational theories: 
\beq
\label{anyonthird}
\canyon_3 \( \frac{\pi}{3} , 1 \) = - 
\canyon_3 \( \frac{2\pi}{3} , -1 \) = \frac{2}{3},
~~~~~~ 
\canyon_3 \( \frac{\pi}{3} , -1 \) = -
 \canyon_3 \( \frac{2\pi}{3} , 1 \) = \frac{8}{9}
\eeq
An interesting feature of this case is that these  gap
parameters can be obtained from an interacting theory with no
statistical chemical potential.  Consider for example the bosonic
theory with $\canyon_3 = 2/3$.  There are two particles with gap parameters
 $z_1 = e^{i\pi/3}, z_2 = e^{-i\pi/3}$. These are the solution to
the bosonic gap equation 
\beq
\label{anyon2}
z_1 = (1-z_1)^{-h_{11}} (1-z_2)^{-h_{12}}, 
~~~~z_2 = (1-z_1)^{-h_{21}} (1-z_2)^{-h_{22}}
\eeq
with real interaction parameters
$h_{11} = h_{22} = 1/2$, $h_{12}=h_{21} = -1/2$. 
This observation suggests that anyons can arise in an ordinary  interacting
theory without inclusion of a Hopf term.  

Another example we have found is for particles of spin $1/4$:
\beq
\label{Anyonfourth} 
\canyon_3 \( \frac{\pi}{2} , -1 \) =
- \canyon_3 \( \frac{\pi}{2} , 1 \) = \frac{3}{16}
\eeq
It remains unknown to us whether there are other values of
$\vtheta$ that correspond to rational theories.

\section{Phase Structure}

In this section we reintroduce the chemical potential and
address a basic issue:  for given interaction parameter $h$
and chemical potential $\mu$, is there a fixed point
with a well-defined $\cfp$?   Generally speaking, $\mu$ is 
related to the density of particles, where $\mu>0$ ($\mu<0$) 
corresponds to increased (decreased) density. 
This issue has largely been ignored even in the 2D integrable
TBA approach.

Given the interaction parameters $h$ and the solutions to the gap
equation $z$, $\cfp$ as given in eq. (\ref{cLr}) is not necessarily
real,  or even well-defined, and this could signify something like a phase 
transition.   In this section we explore this issue for
1-particle theories, where a real $\cfp$ requires a real $z$.  
Of course as the last section shows, real $\cfp$ can arise for
complex $z$ in multi-particle theories. 

The functions $\Lr_{d+1} (z)$ are real for $-\infty < z < 1$.  This implies
the following physical ranges of $z$:
\barray
\nonumber
&~&{\bf bosons:}~~~~~~~ -\infty < z < 1, ~~~~~~ -2\( 1-\inv{2^d} \) < 
\cfp (z) < 1
\\
 &~&{\bf fermions:}~~~~~~~ -1 < z < \infty , ~~~~~~ -1 < 
\cfp (z) <  2\( 1-\inv{2^d} \)
\label{ranges}
\earray
In both cases $\cfp < 0$ corresponds to $z< 0$.   

Let us write the gap equation as 
\beq
\label{fgap}
f^{(h)}_\mp (z) \equiv \frac{z}{( 1\mp z)^{\mp h}} = e^{\mu/T} 
\eeq
where $\mu$ is the chemical potential.  We are only considering
fixed points here,  so $\mu$ must be proportional to $T$ 
for it to have any affect on $\cfp$.     Given $h,\mu$, one can
graphically obtain the gap $z$ by first  plotting the 
function $f^{(h)}_\mp (z)$ as a function of $z$ for the ranges
in eq. (\ref{ranges}).  A given $\mu$, by eq. (\ref{fgap}), is 
a point on the y-axis.  Thus, the intersection of a horizontal
line crossing the $y$-axis at $e^{\mu/T}$ with the $f^{(h)}$ curves
determines the gap $z$.   There are several cases depending on $h$
and whether the particles are bosons or fermions. 

\bigskip
\noindent 
{\bf Bosons with $h<0$.} ~~~This case is shown in figure 2.  One sees
that for any real $\mu$ there is a single solution with $0<z<1$.  
Solutions with a negative $z$, on the negative $y$-axis,
 require the chemical potential to 
have an imaginary part $Im (\mu ) =  \pi T$.  As explained in
the last section, this can correspond to a statistical chemical potential
with $\vtheta = \pi$.  
 In this regime there are two solutions  to $z$
if $h<-1$, one solution if $-1<h<0$, 
and no solution if the real part of $\mu$ is too large.  
Since in all cases a negative $z$ requires an imaginary part
 to the chemical potential, 
throughout this section by ``complex chemical potential'' we mean
one where specifically   $Im (\mu ) = \pi T$. 

\begin{figure}[htb] 
\begin{center}
\hspace{-15mm}
\psfrag{E}{$e^{\mu/T}$}
\psfrag{fm}{$f^{(h)}_-$}
\psfrag{V}{$ {\inv{1+h}}$}
\psfrag{z}{$z$}
\psfrag{1}{$1$}
\includegraphics[width=12cm]{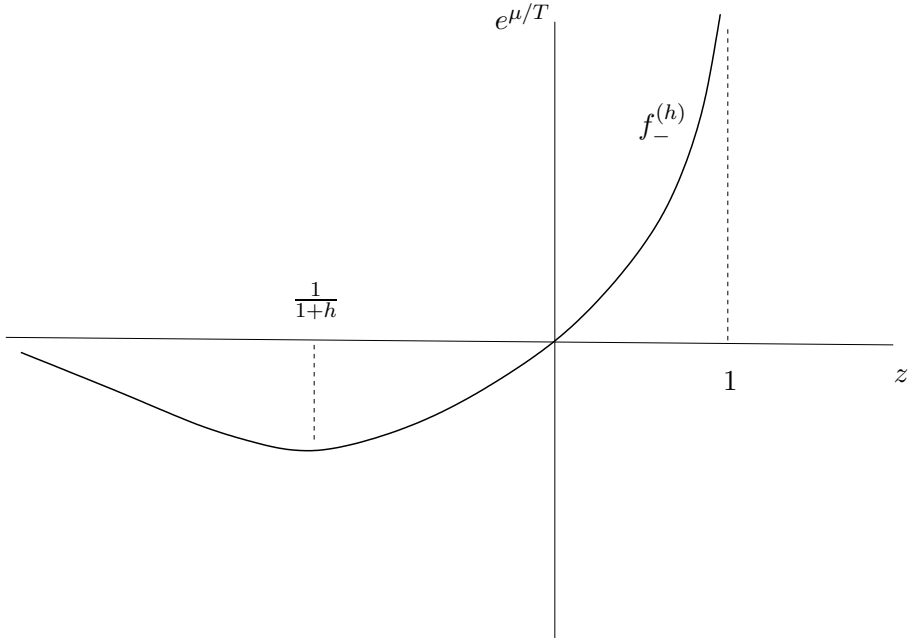} 
\end{center}
\vspace{-2mm}
\caption{Bosons with $h<0$.} 
\vspace{-2mm}
\label{Figure2} 
\end{figure}

\bigskip

\noindent 
{\bf Bosons with $h>0.$} ~~~This case is shown in figure 3.  Here,
for real chemical potential, there is no solution unless $\mu$ is
negative and
\beq
\label{musol1}
\mu/T < - \( (h+1)\log (h+1) - h \log h \) 
\eeq
in which case there are two solutions with $0<z<1$. Thus there is
no solution for zero chemical potential.   The horizontal
line displaying the two solutions is rendered in the figure. 
This implies that if the gas is too dense,  there is no fixed point.
The latter means for instance that the black body formula breaks down.
This could signify a kind of Bose-Einstein condensation. 
  This explains for
example 
why one cannot obtain a well-defined UV limit for the 
bosonic version of the sinh-Gordon model,  as explored in \cite{Mussbos},
unless one lowers the density.  
As the density is decreased, there are two possible $\cfp$.
It is not entirely clear what the significance is of these two
solutions.   If they are both possible in the same model, then this  
suggests  there are  two phases and which one is reached depends on 
the details of the RG flow.   
  On the other hand there
is always a solution with $z<0$ with a complex chemical potential. 
\bigskip

\begin{figure}[htb] 
\begin{center}
\hspace{-15mm}
\psfrag{E}{$e^{\mu/T}$}
\psfrag{fm}{$f^{(h)}_-$}
\psfrag{V}{$ {\inv{1+h}}$}
\psfrag{z}{$z$}
\psfrag{1}{$1$}
\includegraphics[width=12cm]{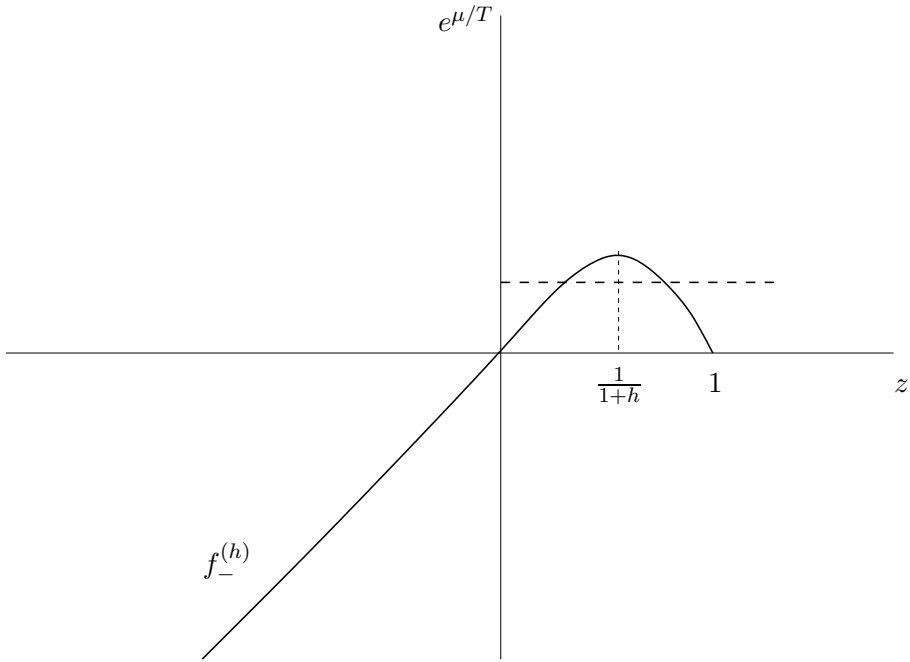} 
\end{center}
\vspace{-2mm}
\caption{Bosons with $h>0$.} 
\vspace{-2mm}
\label{Figure3} 
\end{figure} 

\noindent
{\bf Fermions with $h<0$.}  ~~~See figure 4.   For any real $\mu$ there
is a single solution with $0<z<\infty$.  With complex chemical potential
there are two solutions if $Re (\mu)$ is not too large.

\begin{figure}[htb] 
\begin{center}
\hspace{-15mm}
\psfrag{E}{$e^{\mu/T}$}
\psfrag{fp}{$f^{(h)}_+$}
\psfrag{V}{${\inv{h-1}}$}
\psfrag{z}{$z$}
\psfrag{1}{$\scriptstyle{-1}$}
\includegraphics[width=12cm]{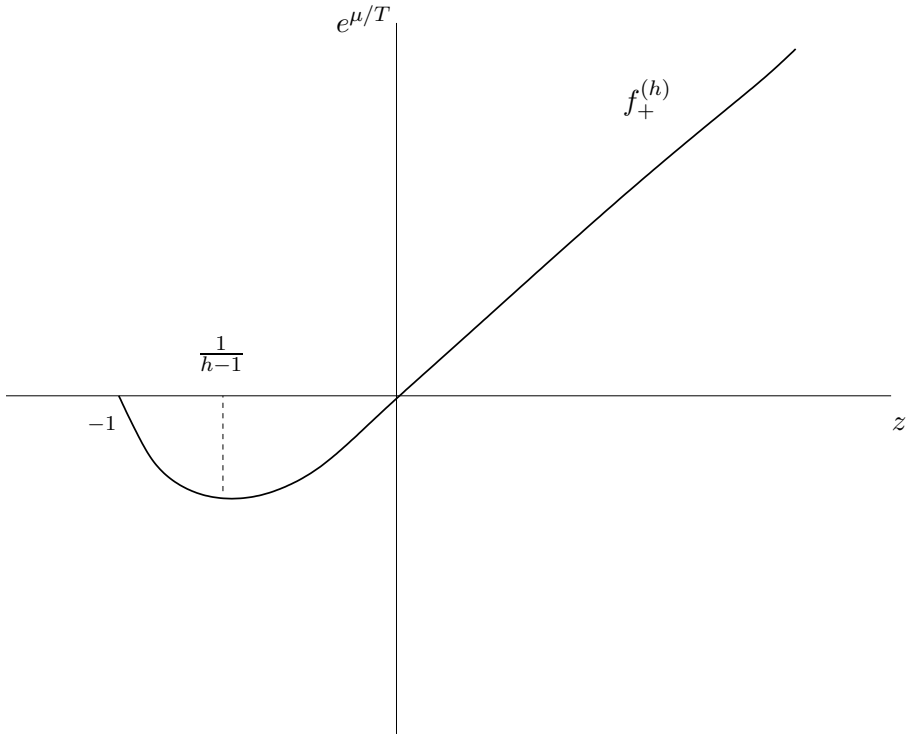} 
\end{center}
\vspace{-2mm}
\caption{Fermions  with $h<0$.} 
\vspace{-2mm}
\label{Figure4} 
\end{figure}

 \noindent
{\bf Fermions with $h>0$.}~~~ See figure 5.   For $0<h<1$,  
and real negative chemical potential there is one solution with $0<z<\infty$.
The sinh-Gordon model is the special case $h=1$ where there
is one solution with $z=\infty$.   
For $h>1$ there is no solution unless 
\beq
\label{mumin}
\mu/T < - \( h \log h - (h-1) \log (h-1) \)
\eeq
in which case there are two solutions.  Finally with complex chemical
potential there is always a unique solution with $-1<z<0$.

\begin{figure}[htb] 
\begin{center}
\hspace{-15mm}
\psfrag{E}{$e^{\mu/T}$}
\psfrag{fp}{$f^{(h)}_+$}
\psfrag{V}{${\inv{h-1}}$}
\psfrag{z}{$z$}
\psfrag{1}{$\scriptstyle{-1}$}
\includegraphics[width=12cm]{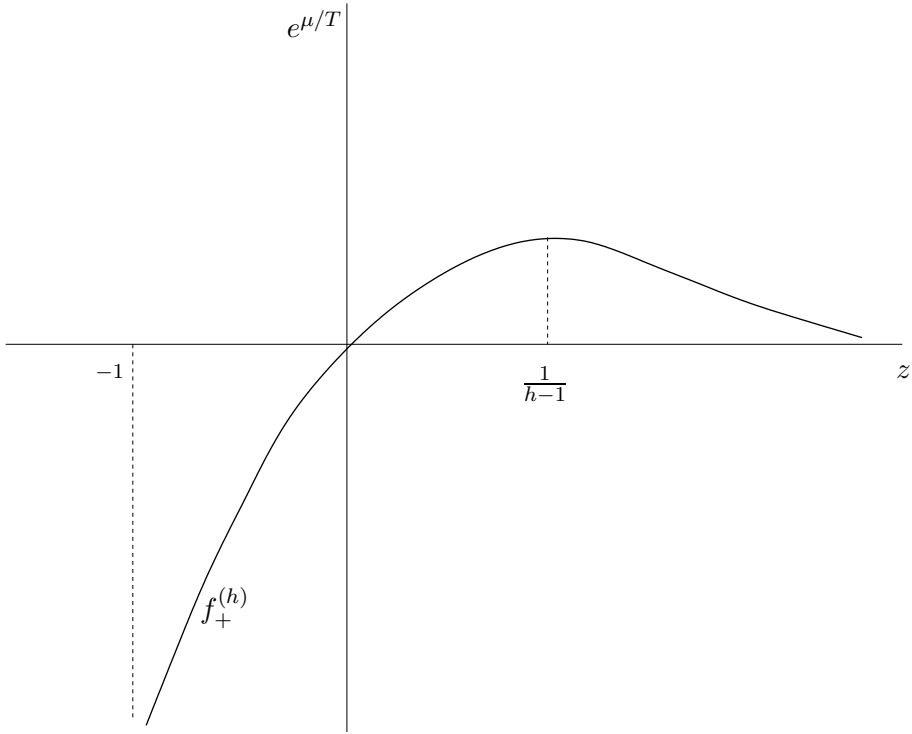} 
\end{center}
\vspace{-2mm}
\caption{Fermions  with $h>0$.} 
\vspace{-2mm}
\label{Figure5} 
\end{figure}

\vfill\eject

\section{Closing  remarks}

The quasi-particle form of the finite temperature partition function
and the integral gap equation is a new formulation of finite
temperature field theory that appears to be well suited
for capturing certain phenomenon in a clear fashion.  
  Though the method is 
an approximation in general, and the nature of this
approximation needs further clarification, 
  we believe that it captures the essential
mechanism that leads to non-trivial fixed points:  interactions
lead to thermal gaps that are proportional to the temperature,
$\Delta = g T$, and $\cfp$ is given in terms of some very special
combinations of polylogarithmic functions eq. (\ref{Lrogers})
of the complex variable
$z= e^{-g}$.    The spectacular properties of these functions
$\Lr_n (z)$  lead us to believe that this is a general property, i.e.
that any non-trivial fixed point has $\cfp$ given by 
the expression (\ref{many6}).   The classification of
rational conformal theories in any dimension 
is then  essentially mapped onto the problem 
of the construction of polylogarithmic ladders in mathematics. 
It would be very interesting to use this idea to explain 
some of the rational values of $\cfp$ obtained for superconformal
gauge theories using string techniques\cite{Gubser}.

The thermal 
gaps are related to the fundamental interaction parameters of the
theory by the algebraic gap equation (\ref{gapz}).  
There may be other methods that lead to the algebraic gap equation 
besides the integral gap equation proposed in section V, 
such as the finite size effective potential\cite{Sachdev}, and this
is important to explore.  

One of the main questions left unanswered in this paper is
how the structure of the stress-tensor form factors can lead
to finite interaction parameters $h_{ab}$ that are RG invariants.
In 2D this is well-understood using integrability and the relation 
of the form factors to the exact S-matrix.   Understanding
this is clearly necessary for further progress on specific models.

We have touched only briefly on the possible applications.  
Further work is clearly needed on some of the suggestions made in 
this paper.  For example, the proposals for the 3D Ising model
and the free energy of anyons in 3D should be compared with 
numerical simulations of the free energy.    

To phrase properties of the equation of state of matter in cosmology
in terms of the c-function $\cqstat (T) $ appears to be fruitful.
As we argued in section IV,  this leads to the bounds 
$-1< w < -1/3$ for dark energy,  but perhaps more importantly
we have related the temperature dependence of $w$ to 
the RG flow of $\cqstat$,  and thus ``c-theorems'' become relevant.   
For instance,  from this perspective 
the fate of the accelerated expansion depends on whether 
the dark energy has an infra-red fixed point.

\section{Acknowledgments}

I wish to thank D. Bradley, J. Cardy, C. Csaki, 
K. Intriligator, G. Mussardo, M. Peskin and  G. Sierra for discussions.

\end{document}